\documentclass[twocolumn,amsmath,prx,superscriptaddress]{revtex4-1}
\usepackage{epsfig,psfrag}
 \usepackage{amsbsy}
\usepackage{amsfonts}
\usepackage{wasysym}
\usepackage{bbm}
\usepackage{tabularx}
\usepackage{euscript}
\usepackage{amsfonts}
\usepackage{exscale}
 \usepackage{slashed}
\usepackage{subfigure}
\usepackage{comment}

\usepackage{amsmath,amssymb,wasysym}
\usepackage{graphicx}
\usepackage{dcolumn}
\usepackage{bm}
\usepackage{braket}
\usepackage{verbatim}
\usepackage{float}
\usepackage{bbold}
\usepackage{color}
\usepackage{xcolor}
\usepackage{relsize}
\usepackage{amsthm}
\usepackage{enumerate}
\usepackage{soul,xcolor}
\usepackage{array}

\usepackage[T1]{fontenc}
\usepackage[colorlinks=true ,urlcolor=blue,urlbordercolor={0 1 1}]{hyperref}

\newcommand{\beq}{\begin{eqnarray}}
\newcommand{\eeq}{\end{eqnarray}}

\newcommand{\bsp}{\begin{split}}
\newcommand{\esp}{\end{split}}

\newcommand{\be}{\begin{equation}}
\newcommand{\ee}{\end{equation}}

\begin{document}

\setstcolor{red}

\title{ XY* transition and extraordinary boundary criticality from fractional exciton condensation in quantum Hall bilayer}
\author{
Ya-Hui Zhang}
\email{yzhan566@jhu.edu}
\affiliation{Department of Physics and Astronomy, Johns Hopkins University, Baltimore, Maryland 21218, USA
}
\author{Zheng Zhu}
\email{zhuzheng@ucas.ac.cn}
\affiliation{Kavli Institute for Theoretical Sciences, University of Chinese Academy of Sciences, Beijing 100190, China}
\affiliation{CAS Center for Excellence in Topological Quantum Computation, University of Chinese Academy of Sciences, Beijing, 100190, China}

\author{Ashvin Vishwanath}
\email{avishwanath@g.harvard.edu}
\affiliation{Department of Physics, Harvard University, Cambridge MA 02138.}
\date{\today}

\begin{abstract}
XY* transitions represent one of the simplest examples of unconventional quantum criticality, in which fractionally charged excitations condense into a superfluid, and display novel features that combine quantum criticality and fractionalization.  Nevertheless their experimental realization is challenging.  Here we propose to study the XY* transition in quantum Hall bilayers at filling  
 $(\nu_1,\nu_2)=(\frac{1}{3},\frac{2}{3})$ where the exciton condensate (EC) phase plays the role of the superfluid. Supported by exact diagonalization  calculation, we argue that there is a continuous transition between an EC phase at small bilayer separation to a pair of 
 decoupled fractional quantum Hall states, at large separation. The transition is driven by condensation of a fractional exciton, a bound state of Laughlin quasiparticle and quasihole,  and is in the XY* universality class.  The fractionalization is manifested by unusual properties including a large anomalous exponent and fractional universal conductivity, which can be conveniently measured through inter-layer tunneling and counter-flow transport, respectively. We also show that the edge is likely to realize the newly predicted extra-ordinary log boundary criticality. Our work highlights the promise of quantum Hall bilayers as an ideal platform for exploring exotic bulk and boundary critical behaviors, that are amenable to immediate experimental exploration in dual-gated bilayer systems. The XY* critical theory can be generalized to a bilayer system with an arbitrary Abelian state in one layer and its particle-hole partner in the other layer. Therefore we anticipate many distinct XY* transitions corresponding to  the different Laughlin states and  Jain sequences in the single layer case.
\end{abstract}

\pacs{Valid PACS appear here}
\maketitle

\section{Introduction} 

The study of quantum phase transitions and universal critical behaviors is one of the major focuses in condensed matter physics\cite{sachdev1999quantum,sondhi1997continuous}. Although many quantum critical points (QCP) are described by the well-established Landau-Ginzburg theory, exceptions arise due to fractionalization beyond the conventional symmetry breaking order. One category is the deconfined quantum critical points(DQCPs) between two different symmetry breaking phases\cite{senthil2004deconfined}. Another category is phase transitions between one phase with fractionalization or topological order and another conventional phase. One simple example is the XY* transition, initially discussed between a Z$_2$ topological ordered insulator (or quantum spin liquid) and a superfluid (or XY ferromagnetism) phase\cite{chubukov1994universal,isakov2012universal,wang2021fractionalized,schuler2022emergent}. The critical theory of  such a  transition is well understood\cite{chubukov1994universal,schuler2022emergent} and its existence in lattice models has been numerically verified\cite{isakov2012universal,wang2021superconductor}. However, experimental observation of the XY* transition is still elusive. Given that even the unambiguous experimental realization of a Z$_2$ spin liquid phase is a great challenge, and that recent progress in synthetic quantum systems target topological order in the absence of global $U(1)$ symmetry \cite{Ruben,Google,Lukin}, the experimental study of an XY* QCP adjacent to a quantum spin liquid phase remains challenging for the near future.

 Here we turn to quantum Hall systems, where fractionalization itself has been well established at fractional fillings\cite{stormer1999fractional}. It is natural to imagine that experimental realization of a QCP with fractionalization in quantum Hall systems is easier, though such a possibility has not been well explored except on plateau transitions\cite{sondhi1997continuous}.  Here we consider the quantum Hall {\em bilayer} system with the electron gases in two layer separated by an insulating barrier, giving rise to  two separate Landau levels  coupled together through the Coulomb repulsion~\cite{Girvin,Halperin1983,Eisenstein2004,Eisenstein2014}.  The fillings in the two layers $\nu_1,\nu_2$ can be controlled separately. In addition, one can tune $d/l_B$ experimentally to study the possible phase transitions. Here $d$ is the distance between the two layers and $l_B$ is the magnetic length.  At small $d/l_B$, it is known that the ground state is an exciton condensation phase~\cite{moon1995spontaneous,yang1996spontaneous,Eisenstein2004,Eisenstein2014,liu2017quantum,li2017excitonic} for the whole line of $\nu_1+\nu_2=1$. There have been many theoretical discussions on other possible phases at larger $d/l_B$ at $(\nu_1,\nu_2)=(\frac{1}{2},\frac{1}{2})$\cite{bonesteel1996gauge,kim2001bilayer,schliemann2001strong,stern2002strong,simon2003coexistence,sheng2003phase,park2004spontaneous,shibata2006ground,moller2008paired,moller2009trial,milovanovic2009nonperturbative,alicea2009interlayer,papic2012disordering,sodemann2017composite,isobe2017interlayer,zhu2017numerical,lian2018wave,wagner2021s}.

 Recently, the evolution under tuning $d/l_B$ was experimentally investigated for this filling $(\nu_1,\nu_2)=(\frac{1}{2},\frac{1}{2})$\cite{liu2022crossover}. There one only finds a crossover between Bose-Einstein-condensation (BEC) regime to Bardeen–Cooper–Schrieffer (BCS)  regime all within a single exciton condensation (EC) phase.  There has been theoretical discussion of superfluid to insulator transition at $(1/2,1/2)$ filling\cite{yang2001dipolar}, but we are not aware of any experimental observation so far.  In contrast, for filling   $(\frac{1}{3},\frac{2}{3})$ (or relatedly $(\nu_1,\nu_2)=(\frac{1}{3},-\frac{1}{3})$, where by $\nu<0$ we mean the system is hole doped relative to the charge neutrality), a phase transition is bound to happen.  At small $d/l_B$ the ground state is still an exciton condensation phase. In the large $d/l_B$ limit, the two layers decouple and the top layer is in the $\nu=\frac{1}{3}$ Laughlin state~\cite{Laughlin83}, while the bottom layer is in the $\nu=-\frac{1}{3}$ (or $\nu=\frac{2}{3}$) Laughlin state. This large $d/l_B$ phase can be viewed as a fractional quantum spin Hall insulator (FQSH)  with a K matrix $K=\begin{pmatrix} 3 & 0 \\ 0 & -3\end{pmatrix}$ if we view the layer as a pseudospin.   Given the recent experimental progress in tuning $d/l_B$ at $(\nu_1,\nu_2)=(\frac{1}{2},\frac{1}{2})$, experimental measurements at filling $(\nu_1,\nu_2)=(\frac{1}{3},-\frac{1}{3})$ or $(\nu_1,\nu_2)=(\frac{1}{3},\frac{2}{3})$) should be straightforward. Note, while conceptually one can think of changing the separation $d$, in experiment  one can tune the ratio $d/l_B$  more conveniently by simultaneously changing the magnetic field and density to keep the filling constant. Thus the transition is well within experimental reach\cite{zeng2021study}!  Actually there already exists some experimental evidence of a direct transition between the exciton superfluid and FQSH phase at filling $(\nu_1,\nu_2)=(\frac{1}{3},\frac{2}{3})$)\cite{champagne2008charge} in GaAs quantum well system. However, the nature of the phase transition is not clear from the existing experimental data. A previous theoretical work already studied similar transition in a model with a  hard-core interaction and suggested the transition is in the XY universality class\cite{chen2012interaction}. Here we will provide numerical evidence for a continuous transition in a realistic model with Coulomb interactions and also point out that this is a XY* transition and propose experimental signatures of the distinction compared to the simple superfluid to Mott insulator transition in the conventional XY universality class. 

We performed exact diagonalization (ED)~\cite{Haldane1985}  for the Coulomb coupled quantum Hall bilayer at filling $(\nu_1,\nu_2)=(\frac{1}{3},\frac{2}{3})$ and found a direct transition between the EC phase at small $d/l_B$ and the FQSH phase at large $d/l_B$. The transition appears continuous in the finite size calculation, suggesting the possibility of a continuous quantum critical point (QCP).  Motivated by the numerical calculation, we propose a critical theory between the EC and FQSH phases in the universality class of XY* transition. Starting from the FQSH phase, the Laughlin electron and Laughlin hole in the two layers bind to form a fractional exciton, with bosonic statistics whose condensation then confines all the anyons and leads to the EC phase at small $d/l_B$. The critical theory is described by the superfluid to insulator transition of the fractional exciton, which carries an exciton charge $1/3$ compared to the ordinary exciton.  We also discuss the realization of an extra-ordinary boundary criticality\cite{metlitski2022boundary} in the edge at this QCP. The XY* transition here can be easily generalized to the case with an arbitrary Abelian FQHE phase in one layer and its particle-hole partner in the other layer. Thus we anticipate many different XY* transitions in the $(D,d/l_B)$ parameter space, where $D$ is the displacement field to tune the exciton density.

\begin{figure}[ht]
\centering 
\includegraphics[width=0.5\textwidth]{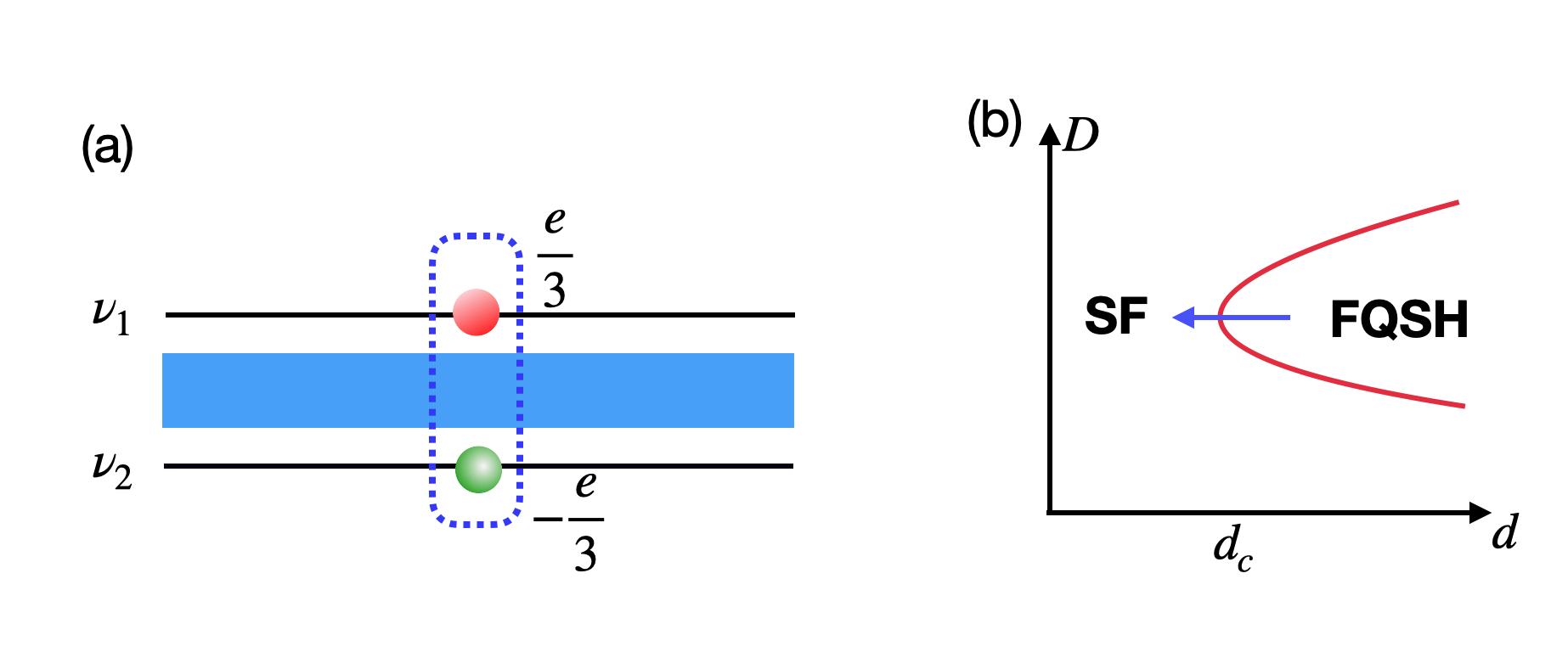}
\caption{(a) Illustration of quantum Hall bilayer, with an insulating layer (blue) in between the two 2DEGs  (b) schematic phase diagram in terms of $(D,\,d)$ while fixing $\nu_1+\nu_2=0$. The displacement field $D$ is the chemical potential for excitons. The distance $d$ tunes inter-layer Coulomb interaction strength. We are interested in the quantum phase transition (indicated by the blue arrow) tuned by $d$ at fixed exciton density $(\nu_1,\nu_2)=(\frac{1}{3},-\frac{1}{3})$. SF denotes the exciton condensation phase with $\langle c^\dagger_1 c_2 \rangle \neq 0$ and FQSH denotes fractional quantum spin Hall insulator formed by two decoupled Laughlin states with opposite chiralities.}
\label{fig:exciton}
\end{figure}

\section{Model and symmetry} 

We consider the quantum Hall bilayer at filling $(\nu_1,\nu_2)=(x,-x)$, illustrated in Fig.~\ref{fig:exciton}.  Here $\nu_a=\frac{N_e}{N_{\Phi}}$, where $N_{\Phi}$ is the number of the magnetic flux in the system. $\nu_2=-x$ means that the system is hole doped with hole density at $x$ per flux.  We will be mainly focused on $x=\frac{1}{3}$, but similar physics can happen for other rational $x$ with an incompressible Abelian FQHE state in the decoupling limit.  $x$ here is the exciton density and can be tuned through the displacement field $D$, while the total filling $\nu_1+\nu_2$ is fixed to be $0$. Up to a stacking of an integer quantum Hall state at the layer $2$, it is also equivalent to consider the filling $(\nu_1,\nu_2)=(x,1-x)$ with $\nu_1+\nu_2=1$. Thus we also consider the filling $(\nu_1,\nu_2)=(\frac{1}{3},\frac{2}{3})$.

We have the Hamiltonian:

\begin{equation}
	\mathcal  H=\frac{1}{2} \sum_{a,b=1,2} V_{ab}(\mathbf q):  \rho_a(\mathbf q) \rho_b(-\mathbf q) :
\end{equation}
where $\rho_a(\mathbf q)=\int d^2\mathbf q \rho_a(\mathbf r) e^{-i \mathbf q \cdot \mathbf r}$ and $\rho_a(\mathbf r)$ is the charge density at layer $a$ projected to the lowest Landau level. We have the Coulomb interaction $V_{11}(q)= V_{22}(q)=\frac{e^2}{\varepsilon q}$ and $V_{12}(q)=V_{21}(q)= \frac{e^2}{\varepsilon q} e^{-qd}$. $d$ represents the distance between two layers in the unit of magnetic length  $l_B=\sqrt{\hbar c/eB}$.

The Hamiltonian considered above has an anti-unitary symmetry $\mathcal M \mathcal C \mathcal T$ for  the quantum Hall bilayer at filling $(\nu_1,\nu_2)=(x,-x)$. The symmetry is a combination of layer exchange symmetry $\mathcal M$, charge conjugation $\mathcal C$ and time reversal $\mathcal T$.  We define electron operators in layer 1 and 2 as $c_1(\mathbf r)$ and $c_2(\mathbf r)$.  The symmetry $\mathcal M \mathcal C \mathcal T$ acts as: $c_1(\mathbf r) \rightarrow c_2^\dagger(\mathbf r), c_2(\mathbf r) \rightarrow c_1^\dagger(\mathbf r)$ combined with complex conjugate $\mathcal K$. Under $\mathcal M \mathcal C \mathcal T$, we have $\rho_1(\mathbf r)\rightarrow -\rho_2(\mathbf r)$ and $\rho_1(\mathbf q)\rightarrow -\rho_2(-\mathbf q)$. One can check that the Hamiltonian satisfies this symmetry.

\begin{figure*}[tbp]
\begin{center}
\includegraphics[width=0.9\textwidth]{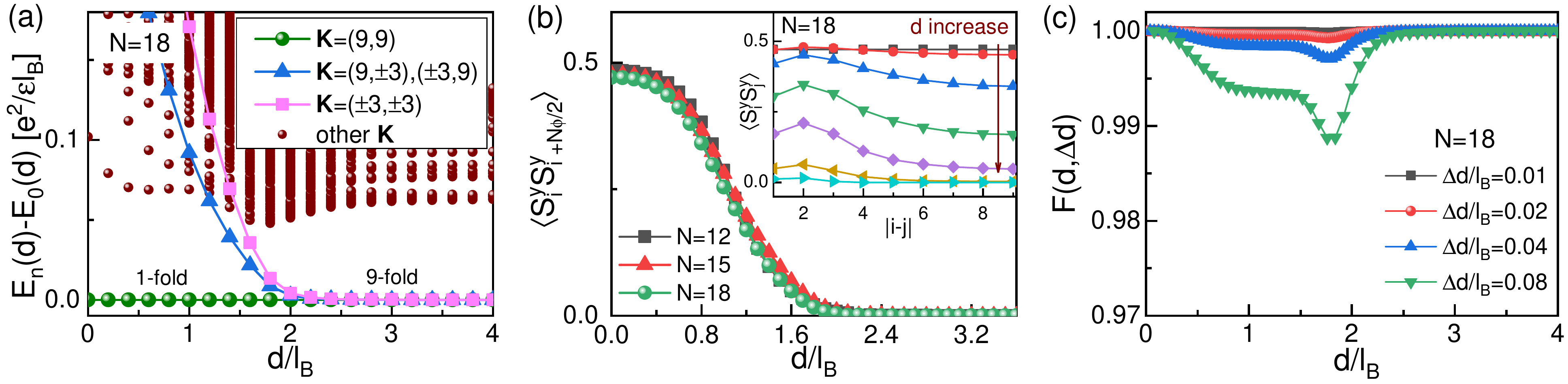}
\end{center}
\par
\renewcommand{\figurename}{FIG.}
\caption{(a) The low-lying energy spectra as a function of layer distance $d/l_B$ for the filling $(\nu_1,\nu_2)=(\frac{1}{3},-\frac{1}{3})$. Here we only show the inequivalent momentum sectors. There are 9-fold degenerate states at larger $d/l_B$ (b) The correlator $\langle {{S^y_i}{S^y_{i+N_\phi/2}}}\rangle $ versus layer distances $d/l_B$, the inset shows $\left\langle {{S^y_i}{S^y_j}} \right\rangle $ as a function of orbital distance $|i-j|$. Here, $i,j=1,\ldots, N_\phi$ are orbital indices and the corresponding distance is $|i-j|L/N_\phi$. From top to bottom, $d/l_B$ ranges from 0 to 2.4 with interval 0.4.  (c) The fidelity  $F(d,\Delta d)$ as a function of the layer distances $d/l_B$ taking different intervals of parameters $\Delta d/l_B$. Note that with decreased $\Delta d$ the dip is weakened.}
\label{Fig:ED}
\end{figure*}


\section{Phase diagram.} 

We fix the filling to be $(\nu_1,\nu_2)=(\frac{1}{3},\frac{2}{3})$ and study the phase diagram of tuning $d/l_B$ through ED. At small $d/l_B$, the system is in an exciton condensation phase with an order parameter $\langle c_1^\dagger c_2 \rangle \neq 0$. Here $c_1,c_2$ are annihilation operator of electron in layer 1 and 2 respectively. At large $d/l_B$, the two layers decouple and we have a fractional quantum spin Hall (FQSH) phase(up to stacking an integer quantum Hall state at layer 2) if viewing layers 1 and 2 as spin up and spin down. The question is whether there is a direct phase transition or an intermediate phase in between. 

Figure~\ref{Fig:ED} (a) shows the flow of low-lying energies with layer distance $d/l_B$. For simplicity, we set $l_B=1$ in the following discussions. We use a torus geometry and the Landau gauge. The evolution of energy spectra indicates a direct transition at $d_c\approx1.7$. When $d>d_c$, we can identify a 9-fold near degeneracy expected for decoupled  $\nu=\pm 1/3$ Laughlin states in the two layers.  When approaching $d_c$ from large $d$, the topological order indicated by the ground state degeneracy disappears at $d_c$.

The phase at $d<d_c$ is an exciton superfluid with order parameter $\langle S^y(\mathbf r) \rangle \neq 0$, where  $S^y(\mathbf r)= i (c^\dagger_{1}(\mathbf r)c_2(\mathbf r)-h.c.)$.  In the lowest Landau level, we have operators $c_{1;m}$ and $c_{2;m}$ where $m$ is the Landau index and labels the position along $x$ direction in our gauge.  So we can define $S^y_m=i(c^\dagger_{1;m}c_{2;m}-h.c.)$, where $m=1,2,..., N_{\Phi}$.  Then we calculate the correlation function $\langle S^y_i S^y_j \rangle$ which is a function of $|i-j|$. In the inset of Fig.~\ref{Fig:ED}(b) we show that $\langle S^y_i S^y_j \rangle$ is almost a constant with $|i-j|$ at small $d$, but decays fast at large $d$.  In particular, we can use $\langle S^y_i S^y_{i+ {N_{\Phi}}/{2}}\rangle$ to characterize the exciton condensation. In Fig.~\ref{Fig:ED}(b) it is clear that $\langle S^y_i S^y_{i+{N_{\Phi}}/{2}}\rangle$ is non-zero at $d<d_c$ and almost vanishes at $d>d_c$.  When approaching $d_c$ from small $d$, the exciton condensation order parameter disappears smoothly across $d_c$.


To further probe the nature of the transition at $d_c$, we compute the ground state fidelity, which is defined by the wave function overlap between the ground state at $d-\Delta d$ and $d$, i.e., $F(d,\Delta d) = \left| {\left\langle \Psi(d-\Delta d)\right|\left. \Psi(d) \right\rangle } \right|$. The fidelity has been shown to be a good indicator to distinguish the continuous transition from the first-order transition for both symmetry-breaking and topological phase transitions~\cite{Zanardi2006,Gu2010}. As shown in Fig.~\ref{Fig:ED} (c), the ground-state fidelity displays a single weak dip at the critical distance $d_c$ instead of showing a sudden jump. Meanwhile, the dip is further weakened with the decrease of $\Delta d$. Thus the numerical evidence indicates the transition might be continuous, though one cannot rule out a weak first-order transition in a finite-size calculation. In the following, we will propose a critical theory for this QCP in the universality class of XY*. The XY* transition is well established to be continuous in other contexts, which further supports the continuous transition scenario of the QCP at $d_c$ from the theoretical side.

\section{Field theory of an XY* transition}

We turn to the filling $(\nu_1,\nu_2)=(\frac{1}{3},-\frac{1}{3})$ for simplicity.  The FQSH phase at the large d limit is described by the following effective field theory:

\begin{equation}
	\mathcal L=-\frac{3}{4\pi}a_1 d a_1+\frac{3}{4\pi}a_2 d a_2+\frac{1}{2\pi}A_1 d a_1-\frac{1}{2\pi}A_2 d a_2
\end{equation}
where $a\,db$ is an abbreviation of $\epsilon_{\mu \nu \sigma} a_\mu \partial_\nu b_\sigma$ . Here, $a_{1;\mu}$ and $a_{2;\mu}$ are emergent dynamical gauge fields while  $A_{1;\mu}$ and $A_{2;\mu}$ are probe fields of the two layers. For example $\vec E_a=-\vec \nabla A_{a;0}-\frac{\partial \vec A_a}{\partial t}$ is the electric field applied to layer $a$. Note in the experiment one can apply $\vec E_1$ and $\vec E_2$ separately and measure currents in a layer resolved fashion.   We then define physical charge $(Q_1,Q_2)$ under $(A_1,A_2)$.  We can also label anyon excitations in terms of their charges $l=(l_1,l_2)$ under $(a_1,a_2)$. The physical charge of the anyon $l$ is $(Q_1,Q_2)=(\frac{1}{3}l_1,\frac{1}{3}l_2)$.  Its statistics is $\theta=\frac{l_1^2-l_2^2}{3}\pi$.   We also make a basis change to define $A_{c;\mu}=\frac{A_{1;\mu}+A_{2;\mu}}{2}$ and $A_{s;\mu}=A_{1;\mu}-A_{2;\mu}$.  The corresponding charge is $Q_c=Q_1+Q_2$ and $Q_s=\frac{Q_1-Q_2}{2}$.  $Q_s$ is the layer pseudospin viewed as a spin $1/2$,  $S_z$  .

The elementary anyon is $l=(\pm 1,0)$ and $l=(0,\pm 1)$ with charge $\pm 1/3$ at each layer.  When we decrease $d$, inter-layer Coulomb interaction increases.  Then an anyon with charge $1/3 $ at layer 1 tends to bind with an anyon with charge $-1/3 $ at layer 2 into an exciton.  When $d$ is further decreased, the binding energy increases and this exciton of anyon can condense and lead to the exciton condensation phase at small $d$. This fractional exciton is labeled by $l=(1,-1)$ with physical charge $(Q_1,Q_2)=(\frac{1}{3},-\frac{1}{3})$ or $Q_c=0, Q_s=\frac{1}{3}$. We label the creation operator of this fractional exciton as $\varphi^\dagger$. The condensation of $\varphi$ is captured by the following critical theory (see the supplementary for derivation):

\begin{align}
	\mathcal L_c&=|(\partial_\mu -i\frac{1}{3}A_{s;\mu})\varphi|^2 -s |\varphi|^2-g|\varphi|^4+\frac{1}{6\pi} A_c d A_s
	\label{eq:critical_theory_third}
\end{align}

When $s<0$, this is a superfluid phase of $A_s$. When $s>0$, we have the correct response of $\frac{1}{6\pi} A_c d A_s$ for the FQSH phase. In principle $\varphi$ is also coupled to a gauge field which however does not affect the critical properties we discuss here due to a Chern Simons term (see Appendix). Note further that tuning the transition at fixed layer density eliminates the single time-derivative chemical potential term. Further, an anti-unitary MCT symmetry (see Appendix) that interchanges layer, and performs the combination of particle hole and time reversal symmetry ensures there is no background flux for the  $\varphi$.   It is clear that the critical theory is the usual `relativistic' XY transition driven by the condensation of a boson which carries charge $1/3$ under $A_s$.   A counter-intuitive feature that is shared with other XY* transitions is that despite the condensation of a fractionally charged boson, the superfluid itself is conventional. One can readily check that the only gauge invariant order parameter is the usual one for integer charge, and all anyons are confined. Alternatively one can show the vortex quantization is the conventional one despite the fractional charge, as a result of attaching an anyon to the fundamental vortex \cite{Kivelson}. In the dual viewpoint, starting from the superfluid phase, the triple vortex becomes gapless and condenses, leading to an insulator. But the elementary vortex of the superfluid phase remains gapped across the QCP and becomes the anyon in the FQSH phase (see the Appendix).

\section{Experimental signatures:} 

We then move to the possible experimental signatures of this unusual QCP. In terms of $\varphi$, Eq.~\ref{eq:critical_theory_third} is the standard critical theory for the XY transition describing interaction tuned superfluid to Mott insulator transition. The critical exponents for thermodynamic quantities are the same as the XY transition. However, the critical boson $\varphi$ here is a non-local field and does not correspond to the microscopic order parameter. Hence the transition is usually called XY* to highlight its difference from the conventional XY transition, which will be manifested in exciton correlation function and conductivity.

\textit{Exciton correlation function:} First, at the QCP, the critical boson has a power law correlation function: $\langle \varphi^\dagger(\mathbf x) \varphi(\mathbf y) \rangle\sim \frac{1}{|\mathbf x-\mathbf y|^{1+\eta}}$ with $\eta \approx 0.038$.  However, the fractional exciton order $\varphi$ is not measurable. The physical order parameter is the conventional exciton operator $\Phi^\dagger=c_1^\dagger c_2$. It is a composite operator in the critical theory: $\Phi=\varphi^3$ and its correlation function has a large decaying exponent: $\langle \Phi^\dagger(\mathbf x) \Phi(\mathbf y) \rangle\sim \frac{1}{|\mathbf x-\mathbf y|^{1+\eta^*}}$ with $\eta^*\approx 3.2$ estimated from the scaling dimension of the $\varphi^3$ mode of the 3D XY universality class.\cite{hasenbusch2011anisotropic}.   The same exponent appears in the correlation function along the time direction, which leads to $\langle \Phi^\dagger(\mathbf r,\omega)\Phi(\mathbf r,-\omega)\rangle\sim \omega^{\eta^*}$ at position $\mathbf r$. 

Interestingly, this exponent can be measured through a local inter-layer tunneling experiment at position $\mathbf r$. Considering a local tunneling term $H'=\Gamma \int d^2 \mathbf r \delta(\mathbf r) c_1^\dagger(\mathbf r)c_2(\mathbf r)+h.c.$, linear response theory derives $I=2e \Gamma^2 \text{Im}\chi_R(\omega=eV,\mathbf r)$ with $I$ and $V$ as current and voltage in the z direction.  $\chi_R(\omega,\mathbf r)$ is the Fourier transformation of $\chi_R(t,\mathbf r)=-i\theta(t)\langle [\Phi^\dagger(\mathbf r,t), \Phi(\mathbf r,0)]\rangle$ in the time direction.\cite{wen1991edge}. So we expect that $\frac{d I}{dV}\sim V^{\eta^*-1}\approx V^{2.2}$ and is  nonlinear to $V$ at zero temperature at $d = d_c$. On the other hand,  when $d<d_c$, we expect $\frac{dI}{dV}$ to have a zero bias peak\cite{eisenstein2014exciton} and when $d>d_c$ it should have a threshold gap. Non-linear I-V curve is expected at the edge of FQHE phase with fractional charge\cite{wen1991edge}. Here we offer an example of the bulk tunneling at the QCP and the large exponent $\eta^*$ is a manifestation of the fractional charge carried by the critical boson. Sometimes it is more convenient to measure a global tunneling\cite{spielman2001observation} from the term $H'=\Gamma \int d^2 \mathbf r  c_1^\dagger(\mathbf r)c_2(\mathbf r)+H.c.$. In this case we expect $I=2e \Gamma^2 \text{Im}\chi_R(\omega=eV,\mathbf q=0)$\cite{stern2001theory}\footnote{In this case, it is important to align the two graphene layers so the inter-layer tunneling probes the spectral weight of the exciton order parameter at momentum $\mathbf q=0$ and energy $\omega=eV$.}.   We have $\frac{dI}{dV}\sim V^{\eta^*-3} \approx V^{0.2}$ at the critical point.

\textit{Universal conductivity:} The XY criticality is known to exhibit a universal conductivity. For our system, we define a $4\times 4$ conductivity tensor in the direct current (DC) limit as $\sigma=\begin{pmatrix} \sigma_{xx} & \sigma_{xy} \\ - \sigma_{xy} & \sigma_{xx} \end{pmatrix}
$, where $\sigma_{xx}$ and $\sigma_{xy}$ are both symmetric $2\times 2$ matrix in layer space. $\sigma_{xx;ab}=\frac{J_{x;a}}{E_{x;b}}$, where $a,b=1,2$ labels the two layers and $\sigma_{xy;ab}$ is defined similarly.

From Eq.~\ref{eq:critical_theory_third} we get the longitudinal conductivity tensor at the QCP to be:
\begin{equation}
	\sigma_{xx}=\frac{\sigma_b}{9}  \frac{e^2}{h}\begin{pmatrix} 1 & -1 \\ -1 &1 \end{pmatrix}
\end{equation}
where $\sigma_b$ is the universal conductivity for the ordinary XY transition, a number of order one in units of $e^2/h$ . The factor of $\frac{1}{9}$ is because the critical boson carries only $1/3$ of the ordinary exciton. Thus the conductivity at this XY* transition is of order $\sim 0.1 \frac{e^2}{h}$. Further, $\sigma_{xy}$ is purely from the background Chern-Simons term in Eq.~\ref{eq:critical_theory_third}. For the filling $(\nu_1,\nu_2)=(\frac{1}{3},-\frac{1}{3})$, we have $	\sigma_{xy}=\begin{pmatrix} \frac{1}{3} & 0 \\ 0 & -\frac{1}{3} \end{pmatrix}\frac{e^2}{h}$.  For $(\nu_1,\nu_2)=(\frac{1}{3},\frac{2}{3})$, we have $	\sigma_{xy}=\begin{pmatrix} \frac{1}{3} & 0 \\ 0 & \frac{2}{3} \end{pmatrix}\frac{e^2}{h}$.

The inverse of  $\sigma$ gives the resistivity tensor: $	\rho=\begin{pmatrix} \rho_{xx} & -\rho_{yx} \\ \rho_{yx} & \rho_{xx} \end{pmatrix}$.  For $(\nu_1,\nu_2)=(\frac{1}{3},-\frac{1}{3})$, we have $\rho_{xx}=\frac{h}{e^2} \sigma_b \begin{pmatrix} 1 &1 \\ 1 & 1 \end{pmatrix}$ and $\rho_{yx}=\frac{h}{e^2}\begin{pmatrix} 3 & 0 \\ 0 & -3 \end{pmatrix}$.   For $(\nu_1,\nu_2)=(\frac{1}{3},\frac{2}{3})$, we have $	\rho_{xx}=\frac{h}{e^2} \frac{1}{4+\sigma_b^2}\begin{pmatrix} 4 \sigma_b & -2\sigma_b \\ -2\sigma_b & \sigma_b\end{pmatrix}
$ and  $\rho_{yx}=\frac{h}{e^2} \frac{1}{4+\sigma_b^2}\begin{pmatrix} 12+\sigma_b^2 & \sigma_b^2 \\ \sigma_b^2 & 6+\sigma_b^2 \end{pmatrix}
$.

The above discussions are exactly at the QCP and zero temperature. In practice, the experiments are always at finite temperature and one expects  critical scaling $\rho(T,\delta)=F(\frac{T}{\delta^{\nu z}})$, where $F$ is a universal function and $\delta=d-d_c$ is the deviation from the critical point. We have $z=1$ and $\nu \approx 0.67$ as the known critical exponents for the XY transition. From collapsing the data of $(T,d-d_c)$ one can extrapolate the exponent $\nu z$ and the universal conductivity. Such a scaling has been performed for the superconductor to insulator transitions\cite{hebard1990magnetic,yazdani1995superconducting,markovic1998thickness,bielejec2002field,wang2021superconductor,hen2021superconductor}.

\begin{figure}[ht]
\centering 
\includegraphics[width=0.5\textwidth]{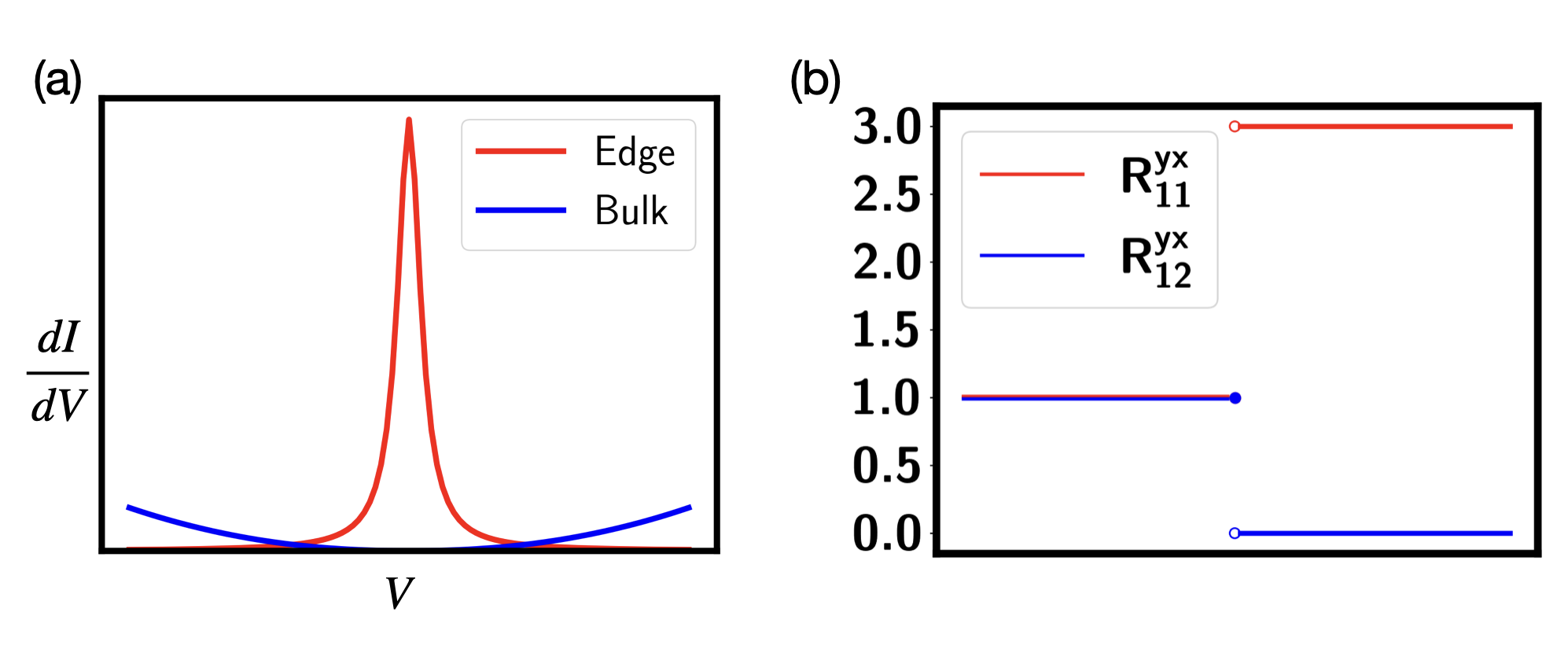}
\caption{(a) Illustration of the different behaviors of inter-layer tunneling I-V curves between the bulk and edge at the critical point. (b) Hall resistivity $R^{xy}_{11}$ and $R^{xy}_{21}$ (at units of $\frac{h}{e^2}$)  in the Hall bar geometry as tuning the distance $d$ for the filling $(\nu_1,\nu_2)=(\frac{1}{3},\frac{2}{3})$. At $d_c$, because of the extra-ordinary boundary criticality, the exciton still behaves like a superfluid at the edge, which guarantees that $R^{yx}_{11}=R^{yx}_{12}=1$. This is very different from the bulk values $R^{yx}_{11}=\frac{12+\sigma_b^2}{4+\sigma_b^2}$ and $R^{yx}_{12}=\frac{\sigma_b^2}{4+\sigma_b^2}$ at the QCP, which are at intermediate values between the $d<d_c$ and $d>d_c$ phases.}
\label{fig:bulk_edge}
\end{figure}

\section{Extra-ordinary log boundary criticality:} 

For the FQSH phase at $d>d_c$, there are helical edge modes.  At $d<d_c$ the helical edge modes will be gapped out by the long-range exciton order. Here let us decide the fate of these edge modes at the critical point. The edge theory of the FQSH phase is:

\begin{equation}
	S_{0}=\int dt dx \frac{1}{2\pi  \tilde \upsilon_F \lambda}\big((\partial_t \tilde \theta)^2-\tilde \upsilon_F^2(\partial_x \tilde \theta)^2 \big)
\end{equation}
where $\tilde \theta$ represents the helical edge modes of the FQSH phase.  Here $e^{i\tilde \theta}$ creates a fractional exciton with charge $1/3$ under $A_s$ at the edge. $\lambda=\frac{4}{3}$ in the decoupling limit and becomes smaller when including inter-layer repulsing at finite $d$. So we have $\lambda<\frac{4}{3}$. At the QCP, it is further coupled to the bulk critical boson through:

\begin{equation}
	S_{\text{boundary}}=S_0-s \int dx dt (e^{i\tilde \theta} \varphi^*+e^{-i \tilde \theta}\varphi)
\end{equation}

We assume the anti-unitary layer exchange symmetry $\mathcal M \mathcal C \mathcal T$ to guarantee that $e^{i\tilde \theta}$ carries zero momentum. Without the $\mathcal M \mathcal C \mathcal T$ symmetry, the above term is absent due to momentum mismatch and disorder needs to be involved, which we leave to future analysis. 

The scaling dimension of the coupling $s$ is $[s]=2-\Delta_{\varphi}-\frac{1}{4} \lambda \approx 0.78-\frac{1}{4}\lambda>0$, where we use $\Delta_\phi=1.22$ as the boundary scaling dimension of the order parameter. So the coupling is relevant and flows to infinity (see the supplementary).  It is thus very likely that it flows to the extraordinary-log boundary critical point\cite{metlitski2022boundary} recently proposed for the 3D XY transition. At this fixed point, the exciton order is almost long-range ordered at the edge: $\langle \Phi^\dagger(x) \Phi(y) \rangle \sim \frac{1}{\log(|x-y|)^{\tilde q}}$. This is in contrast to the large power-law decaying exponent for the correlation function in the bulk, manifested in the inter-layer tunneling $I-V$ curves as illustrated in Fig.~\ref{fig:bulk_edge}(a). Besides, the exciton transport at the edge is still superfluid like with an infinite conductance $G=\frac{1}{9\lambda}$\cite{giamarchi2003quantum} which should be infinite at zero temperature dramatically different from the metallic bulk transport. As a result, transport measurements in Hall bar geometry with edge and in the Corbino geometry without edge are very different at the QCP.  The flow of $\frac{1}{\lambda}$  to zero is only logarithmic, so at finite temperature we expect $G\sim \log \frac{1}{T}$, which may be tested in experiment.

So far we have focused on the filling $(\nu_1,\nu_2)=(\frac{1}{3},-\frac{1}{3})$. For the filling $(\nu_1,\nu_2)=(\frac{1}{3},\frac{2}{3})$, the bulk behavior is exactly the same. But there is an additional edge mode from the integer quantum Hall effect.  At clean sample or weak disorder regime, this integer quantum Hall edge model can not be hybridized with the FQSH edge modes and can be ignored. Thus we expect the same extraordinary critical behavior. For example, at filling $(\nu_1,\nu_2)=(\frac{1}{3},\frac{2}{3})$, we have shown that $\rho^{yx}_{11}$ and $\rho^{yx}_{21}$ in the bulk are at certain fractional values depending on the universal conductivity $\sigma_b$. However, because of the extra-ordinary boundary behavior, we expect $\rho^{yx}_{11}=\rho^{yx}_{21}=1$ in the Hall bar measurement, as illustrated in Fig.~\ref{fig:bulk_edge}(b).  Distinction between edge and bulk transport can be a direct verification of the proposed extra-ordinary boundary criticality. If there is strong disorder, then the two edge modes in the layer 2 may be coupled together and flow to the Kane-Fisher-Polchinski fixed point\cite{kane1994randomness} for the filling $(\nu_1,\nu_2)=(\frac{1}{3},\frac{2}{3})$. If this happens, we expect the coupling of the bulk exciton order parameter to the edge to be irrelevant and we have an ordinary boundary critical behavior\cite{metlitski2022boundary}. It is interesting to study the transition between extra-ordinary boundary criticality and the ordinary boundary criticality tuned by the disorder strength, which we leave to future work.

\section{ XY* transition for General Abelian states}

In the previous sections we focus on the filling $(\nu_1,\nu_2)=(\frac{1}{3},-\frac{1}{3})$.  Here we point out that XY* transition exist for any rational filling $(\nu_1,\nu_2)=(x,-x)$ as long as there is an Abelian FQHE phase at the filling $x$.

Let us consider bilayer with an arbitrary Abelian FQHE state in one layer and its particle-hole partner in the other layer. We still have the  $\mathcal M \mathcal C \mathcal T$ symmetry. Any Abelian FQHE phase can be captured by a $K$ matrix with dimension $N$.  The Low energy theory in the decoupled limit is:

\begin{align}
    \mathcal L&=-\frac{1}{4\pi} \mathbf a_1^T \mathbf K d\mathbf a_1+\frac{1}{4\pi} \mathbf a_2^T \mathbf K d\mathbf a_2 \notag \\ 
    &~~~+\frac{1}{2\pi}  A_1 \mathbf q^T d \mathbf a_1 -\frac{1}{2\pi}  A_2 \mathbf q^T d \mathbf a_2
\end{align}
where $\mathbf K$ is a $N \times N$ matrix. $\mathbf q$ is a $N\times 1$ vector.  Similarly $\mathbf a_1, \mathbf a_2$ are emergent U(1) gauge fields with $N$ components in the two layers.  As before $A_1, A_2$ are probe fields in the two layers with only one component. An Abelian FQHE phase is specified by $\mathbf K$ and $\mathbf q$.  The analysis below applies to any Abelian FQHE phase.

The $\mathcal M \mathcal C \mathcal T$ transforms in the following way: $\big(\mathbf a^0_1(t,\mathbf r),\vec{\mathbf a}_1(t,\mathbf r)\big) \rightarrow \big(-\mathbf a^0_2(-t,\mathbf r),\vec{ \mathbf a}_2(-t,\mathbf r)\big)$, $\big( A^0_1(t,\mathbf r),\vec{A}_1(t,\mathbf r) \big)\rightarrow \big( -A^0_2(-t,\mathbf r), \vec A_2(-t,\mathbf r)\big)$.

Then we redefine $A_c=\frac{1}{2}(A_1+A_2), A_s=A_1-A_2$,  $\mathbf a_c=\mathbf a_1+\mathbf a_2$, $\mathbf a_s=\mathbf a_1-\mathbf a_2$, the action for the decoupled phase is

\begin{align}
    \mathcal L&=-\frac{1}{4\pi} \mathbf a_1^T \mathbf K d\mathbf a_1+\frac{1}{4\pi} \mathbf a_2^T \mathbf K d\mathbf a_2 \notag \\ 
    &~~~+\frac{1}{2\pi} A_c \mathbf q^T d \mathbf a_s+\frac{1}{4\pi}A_s \mathbf q^T d \mathbf a_c
\end{align}

Suppose the lowest charged excitation at each layer is generated by the vector $\mathbf l_0$. Then $\mathbf l=(\mathbf l_0, -\mathbf l_0)^T$ generates a boson with $Q_c=0$ and $Q_s=\mathbf q^T \mathbf K^{-1} \mathbf l_0$.   Let us label this boson as $\varphi$, then the critical theory is

\begin{align}
	\mathcal L&=|(\partial_\mu -i\mathbf l_0^T(\mathbf a_{1;\mu}-\mathbf a_{2;\mu}))\varphi|^2- s |\varphi|^2-g|\varphi|^4 \notag \\ 
 &~~~-\frac{1}{4\pi} \mathbf a_1^T \mathbf K d\mathbf a_1+\frac{1}{4\pi} \mathbf a_2^T \mathbf K d\mathbf a_2 \notag \\ 
 &~~~+\frac{1}{2\pi} A_c \mathbf q^T d \mathbf a_s+\frac{1}{4\pi}A_s \mathbf q^T d \mathbf a_c
\end{align}
which can be rewritten as:

\begin{align}
	\mathcal L&=|(\partial_\mu -i\mathbf l_0^T \mathbf a_{s;\mu}))\varphi|^2- s |\varphi|^2-g|\varphi|^4 \notag \\ 
 &~~~-\frac{1}{4\pi}\mathbf a_c^T \mathbf K d \mathbf a_s+\frac{1}{2\pi} A_c \mathbf q^T d \mathbf a_s+\frac{1}{4\pi}A_s \mathbf q^T d \mathbf a_c
\end{align}

With the assumption that $\det \mathbf K\neq 0$, we can integrate $\mathbf a_c$, which locks $\mathbf a_s=\mathbf K^{-1} \mathbf q A_s$. Then the final critical theory is:

\begin{align}
	\mathcal L&=|(\partial_\mu -iQ_s A_{s;\mu}))\varphi|^2- s |\varphi|^2-g|\varphi|^4+\frac{\sigma^{cs}_{xy}}{2\pi} A_c d A_s
\end{align}
where $Q_s=\mathbf q^T \mathbf K^{-1} \mathbf l_0$.   $\sigma^{cs}_{xy}=\mathbf q^T \mathbf K^{-1} \mathbf q$.

The above action clearly describes a XY* transition with a condensation of a fractional exciton of exciton charge $Q_s=\mathbf q^T \mathbf K^{-1} \mathbf l_0$. Similar to our discussion for $(\nu_1,\nu_2)=(\frac{1}{3},-\frac{1}{3})$, there is a fractional counter-flow conductivity: 

\begin{equation}
    \sigma_{xx}=Q_s^2 \sigma_b \frac{e^2}{h}\begin{pmatrix} 1 & 1 \\ 1 & 1 \end{pmatrix}
\end{equation}
where $\sigma_b$ is again the universal conductivity of the usual XY transition.

One simple example is $(\nu_1,\nu_2)=(\frac{1}{5},-\frac{1}{5})$. Then the K matrix is one dimensional with $K=5$, $l=1$, $q=1$.   We simply reach that $Q_s=\frac{1}{5}$, indicating a fractional exciton with $1/5$ exciton charge at the XY* transition.  A more nontrivial example is to consider $(\nu_1,\nu_2)=(\frac{2}{5},-\frac{2}{5})$ or equivalently $(\frac{2}{5},\frac{3}{5})$. Now in the decoupled phase we should use $K=\begin{pmatrix} 3 & 1 \\ 1 &2 \end{pmatrix}$ and $q=(1,0)^T$.  The smallest charged anyon is generated by $\mathbf l_0=(1,1)^T$, with charge $1/5$ and statistics $\frac{3}{5} \pi$.   In the bilayer setup $\mathbf l=(\mathbf l_0, -\mathbf l_0)^T$ generates a bosonic fractional exciton with charge $Q_s=\frac{1}{5}$. Again we expect a XY* transition with $Q_s=\frac{1}{5}$. In both these cases the universal counter-flow conductivity is $\frac{1}{25}\sigma_b e^2/h $. The physical exciton order parameter is $\Phi=\varphi^5$ and should have an even larger anomalous scaling dimension than the $(\nu_1,\nu_2)=(\frac{1}{3},\frac{2}{3})$ case. For this case a finite inter-layer tunneling destroys the superfluid phase, but the XY* QCP is stable because $-\varphi^5-H.c.$ is known to be irrelevant for the XY transition. In contrast, for the filling $(\nu_1,\nu_2)=(\frac{1}{3},\frac{2}{3})$, an inter-layer tunneling term acts as $-\varphi^3-H.c.$ at low energy and drives the QCP to be first order transition. 

One direct evidence of the fractional exciton in the XY* transition is a fractional universal conductivity. However, this requires the value of $\sigma_b$, the universal conductivity of the ordinary XY transition. Unfortunately there is no measurement or accurate prediction of the universal conductivity $\sigma_b$ so far.  However, in quantum Hall bilayer we can find XY* transitions at different fillings $(x,-x)$ to independently measure both $\sigma_b$ and $Q_s$.  For example, there are a XY* transition at filling $(\nu_1,\nu_2)=(\frac{1}{3},-\frac{1}{3})$ with $Q_s=\frac{1}{3}$, and another XY* transition at filling $(\nu_1,\nu_2)=(\frac{2}{5},-\frac{2}{5})$ with $Q_s=\frac{1}{5}$.  A clear prediction is that the ratio of the universal counter-flow conductivity at the QCP of these two fillings should be $\frac{25}{9}$. There should be many different XY* transitions corresponding to different rational fillings $x=\frac{m}{2pm \pm 1}$ with the charge of the elementary anyon known from well-established theory, thus one can even do scaling between the counter-flow conductivity at the QCP and the expected value of $Q_s$ to test the picture that the critical exciton boson is formed by a pair of anyons.

\section{Summary} 

In conclusion, we proposed a route to accessing the XY* quantum critical point (QCP) by tuning magnetic field (and hence the $d/l_B$ ratio) while keeping the filling fixed at  $(\nu_1,\nu_2)=(\frac{1}{3},-\frac{1}{3})$ or $(\nu_1,\nu_2)=(\frac{1}{3},\frac{2}{3})$ in a quantum Hall bilayer. At such a QCP, two anyons from the FQSH phase in the $d>d_c$ side form a fractional exciton and condense, leading to the exciton condensation phase in the $d<d_c$ side. The fractional charge of the exciton is manifested in the large anomalous exponent of the exciton correlation function and a $1/9$ factor in the universal conductivity. We also argue that the edge at this XY* transition shows extra-ordinary boundary criticality behavior. At criticality, the exciton behaves like a superfluid at the edge despite it showing metallic transport in the bulk. The generalization of our theory to other fillings like $(\nu_1,\nu_2)=(\frac{1}{5},-\frac{1}{5})$ is straightforward. Interestingly, in that case the presence of weak interlayer tunneling, which is expected to introduce a $(\lambda \varphi^5 +{\rm h.c.})$ anisotropy which is expected to be dangerously irrelevant \cite{Sandvik}, which will remove the Goldstone mode in the exciton condensed phase, but will  preserve the critical properties. The relation between anyons in the topological phase and vorticity on the ordered side has led to the proposal of interesting `memory' effects in other contexts \cite{SenthilFisher} which can also be explored here. These directions will be worth exploring in the future. In summary, our work suggests a new approach to studying exotic quantum phase transitions with fractionalization and unusual boundary critical behaviors in the highly tunable quantum Hall bilayer systems.

\textit{Note added} After our manuscript appeared, we became aware of another preprint\cite{wu2023continuous} discussing continuous quantum phase transition in quantum Hall bilayer system, but in a different setup with different physics.

 \textbf{Acknowledgement}  We thank Max Metlitski for insightful comments on the manuscript. YHZ thanks Cory Dean, Jia Li and Yihang Zeng for discussions on related experiments in quantum Hall bilayer. Z.Z. thanks Zlatko Papi\'{c} and Songyang Pu for helpful discussions.  YHZ is supported by a startup fund from Johns Hopkins university.  Z.Z. was supported by the National Natural Science Foundation of China (Grant No.12074375),the Fundamental Research Funds for the Central Universities, the Strategic Priority Research Program of CAS (Grant No.XDB33000000). AV was supported by the Simons Collaboration on Ultra-Quantum Matter, which is a grant from the Simons Foundation (651440, AV) and from NSF-DMR 2220703.

\bibliographystyle{apsrev4-1}
\bibliography{main}

\begin{thebibliography}{69}%
\makeatletter
\providecommand \@ifxundefined [1]{%
 \@ifx{#1\undefined}
}%
\providecommand \@ifnum [1]{%
 \ifnum #1\expandafter \@firstoftwo
 \else \expandafter \@secondoftwo
 \fi
}%
\providecommand \@ifx [1]{%
 \ifx #1\expandafter \@firstoftwo
 \else \expandafter \@secondoftwo
 \fi
}%
\providecommand \natexlab [1]{#1}%
\providecommand \enquote  [1]{``#1''}%
\providecommand \bibnamefont  [1]{#1}%
\providecommand \bibfnamefont [1]{#1}%
\providecommand \citenamefont [1]{#1}%
\providecommand \href@noop [0]{\@secondoftwo}%
\providecommand \href [0]{\begingroup \@sanitize@url \@href}%
\providecommand \@href[1]{\@@startlink{#1}\@@href}%
\providecommand \@@href[1]{\endgroup#1\@@endlink}%
\providecommand \@sanitize@url [0]{\catcode `\\12\catcode `\$12\catcode
  `\&12\catcode `\#12\catcode `\^12\catcode `\_12\catcode `\%12\relax}%
\providecommand \@@startlink[1]{}%
\providecommand \@@endlink[0]{}%
\providecommand \url  [0]{\begingroup\@sanitize@url \@url }%
\providecommand \@url [1]{\endgroup\@href {#1}{\urlprefix }}%
\providecommand \urlprefix  [0]{URL }%
\providecommand \Eprint [0]{\href }%
\providecommand \doibase [0]{http://dx.doi.org/}%
\providecommand \selectlanguage [0]{\@gobble}%
\providecommand \bibinfo  [0]{\@secondoftwo}%
\providecommand \bibfield  [0]{\@secondoftwo}%
\providecommand \translation [1]{[#1]}%
\providecommand \BibitemOpen [0]{}%
\providecommand \bibitemStop [0]{}%
\providecommand \bibitemNoStop [0]{.\EOS\space}%
\providecommand \EOS [0]{\spacefactor3000\relax}%
\providecommand \BibitemShut  [1]{\csname bibitem#1\endcsname}%
\let\auto@bib@innerbib\@empty
\bibitem [{\citenamefont {Sachdev}(1999)}]{sachdev1999quantum}%
  \BibitemOpen
  \bibfield  {author} {\bibinfo {author} {\bibfnamefont {S.}~\bibnamefont
  {Sachdev}},\ }\href@noop {} {\bibfield  {journal} {\bibinfo  {journal}
  {Physics world}\ }\textbf {\bibinfo {volume} {12}},\ \bibinfo {pages} {33}
  (\bibinfo {year} {1999})}\BibitemShut {NoStop}%
\bibitem [{\citenamefont {Sondhi}\ \emph {et~al.}(1997)\citenamefont {Sondhi},
  \citenamefont {Girvin}, \citenamefont {Carini},\ and\ \citenamefont
  {Shahar}}]{sondhi1997continuous}%
  \BibitemOpen
  \bibfield  {author} {\bibinfo {author} {\bibfnamefont {S.~L.}\ \bibnamefont
  {Sondhi}}, \bibinfo {author} {\bibfnamefont {S.}~\bibnamefont {Girvin}},
  \bibinfo {author} {\bibfnamefont {J.}~\bibnamefont {Carini}}, \ and\ \bibinfo
  {author} {\bibfnamefont {D.}~\bibnamefont {Shahar}},\ }\href@noop {}
  {\bibfield  {journal} {\bibinfo  {journal} {Reviews of modern physics}\
  }\textbf {\bibinfo {volume} {69}},\ \bibinfo {pages} {315} (\bibinfo {year}
  {1997})}\BibitemShut {NoStop}%
\bibitem [{\citenamefont {Senthil}\ \emph {et~al.}(2004)\citenamefont
  {Senthil}, \citenamefont {Vishwanath}, \citenamefont {Balents}, \citenamefont
  {Sachdev},\ and\ \citenamefont {Fisher}}]{senthil2004deconfined}%
  \BibitemOpen
  \bibfield  {author} {\bibinfo {author} {\bibfnamefont {T.}~\bibnamefont
  {Senthil}}, \bibinfo {author} {\bibfnamefont {A.}~\bibnamefont {Vishwanath}},
  \bibinfo {author} {\bibfnamefont {L.}~\bibnamefont {Balents}}, \bibinfo
  {author} {\bibfnamefont {S.}~\bibnamefont {Sachdev}}, \ and\ \bibinfo
  {author} {\bibfnamefont {M.~P.}\ \bibnamefont {Fisher}},\ }\href@noop {}
  {\bibfield  {journal} {\bibinfo  {journal} {Science}\ }\textbf {\bibinfo
  {volume} {303}},\ \bibinfo {pages} {1490} (\bibinfo {year}
  {2004})}\BibitemShut {NoStop}%
\bibitem [{\citenamefont {Chubukov}\ \emph {et~al.}(1994)\citenamefont
  {Chubukov}, \citenamefont {Senthil},\ and\ \citenamefont
  {Sachdev}}]{chubukov1994universal}%
  \BibitemOpen
  \bibfield  {author} {\bibinfo {author} {\bibfnamefont {A.~V.}\ \bibnamefont
  {Chubukov}}, \bibinfo {author} {\bibfnamefont {T.}~\bibnamefont {Senthil}}, \
  and\ \bibinfo {author} {\bibfnamefont {S.}~\bibnamefont {Sachdev}},\
  }\href@noop {} {\bibfield  {journal} {\bibinfo  {journal} {Physical review
  letters}\ }\textbf {\bibinfo {volume} {72}},\ \bibinfo {pages} {2089}
  (\bibinfo {year} {1994})}\BibitemShut {NoStop}%
\bibitem [{\citenamefont {Isakov}\ \emph {et~al.}(2012)\citenamefont {Isakov},
  \citenamefont {Melko},\ and\ \citenamefont {Hastings}}]{isakov2012universal}%
  \BibitemOpen
  \bibfield  {author} {\bibinfo {author} {\bibfnamefont {S.~V.}\ \bibnamefont
  {Isakov}}, \bibinfo {author} {\bibfnamefont {R.~G.}\ \bibnamefont {Melko}}, \
  and\ \bibinfo {author} {\bibfnamefont {M.~B.}\ \bibnamefont {Hastings}},\
  }\href@noop {} {\bibfield  {journal} {\bibinfo  {journal} {Science}\ }\textbf
  {\bibinfo {volume} {335}},\ \bibinfo {pages} {193} (\bibinfo {year}
  {2012})}\BibitemShut {NoStop}%
\bibitem [{\citenamefont {Wang}\ \emph
  {et~al.}(2021{\natexlab{a}})\citenamefont {Wang}, \citenamefont {Cheng},
  \citenamefont {Witczak-Krempa},\ and\ \citenamefont
  {Meng}}]{wang2021fractionalized}%
  \BibitemOpen
  \bibfield  {author} {\bibinfo {author} {\bibfnamefont {Y.-C.}\ \bibnamefont
  {Wang}}, \bibinfo {author} {\bibfnamefont {M.}~\bibnamefont {Cheng}},
  \bibinfo {author} {\bibfnamefont {W.}~\bibnamefont {Witczak-Krempa}}, \ and\
  \bibinfo {author} {\bibfnamefont {Z.~Y.}\ \bibnamefont {Meng}},\ }\href@noop
  {} {\bibfield  {journal} {\bibinfo  {journal} {Nature communications}\
  }\textbf {\bibinfo {volume} {12}},\ \bibinfo {pages} {1} (\bibinfo {year}
  {2021}{\natexlab{a}})}\BibitemShut {NoStop}%
\bibitem [{\citenamefont {Schuler}\ \emph {et~al.}(2022)\citenamefont
  {Schuler}, \citenamefont {Henry}, \citenamefont {Lu},\ and\ \citenamefont
  {L{\"a}uchli}}]{schuler2022emergent}%
  \BibitemOpen
  \bibfield  {author} {\bibinfo {author} {\bibfnamefont {M.}~\bibnamefont
  {Schuler}}, \bibinfo {author} {\bibfnamefont {L.-P.}\ \bibnamefont {Henry}},
  \bibinfo {author} {\bibfnamefont {Y.-M.}\ \bibnamefont {Lu}}, \ and\ \bibinfo
  {author} {\bibfnamefont {A.~M.}\ \bibnamefont {L{\"a}uchli}},\ }\href@noop {}
  {\bibfield  {journal} {\bibinfo  {journal} {arXiv preprint arXiv:2204.03659}\
  } (\bibinfo {year} {2022})}\BibitemShut {NoStop}%
\bibitem [{\citenamefont {Wang}\ \emph
  {et~al.}(2021{\natexlab{b}})\citenamefont {Wang}, \citenamefont {Biscaras},
  \citenamefont {Erb},\ and\ \citenamefont {Shukla}}]{wang2021superconductor}%
  \BibitemOpen
  \bibfield  {author} {\bibinfo {author} {\bibfnamefont {F.}~\bibnamefont
  {Wang}}, \bibinfo {author} {\bibfnamefont {J.}~\bibnamefont {Biscaras}},
  \bibinfo {author} {\bibfnamefont {A.}~\bibnamefont {Erb}}, \ and\ \bibinfo
  {author} {\bibfnamefont {A.}~\bibnamefont {Shukla}},\ }\href@noop {}
  {\bibfield  {journal} {\bibinfo  {journal} {Nature communications}\ }\textbf
  {\bibinfo {volume} {12}},\ \bibinfo {pages} {1} (\bibinfo {year}
  {2021}{\natexlab{b}})}\BibitemShut {NoStop}%
\bibitem [{\citenamefont {Verresen}\ \emph {et~al.}(2021)\citenamefont
  {Verresen}, \citenamefont {Lukin},\ and\ \citenamefont {Vishwanath}}]{Ruben}%
  \BibitemOpen
  \bibfield  {author} {\bibinfo {author} {\bibfnamefont {R.}~\bibnamefont
  {Verresen}}, \bibinfo {author} {\bibfnamefont {M.~D.}\ \bibnamefont {Lukin}},
  \ and\ \bibinfo {author} {\bibfnamefont {A.}~\bibnamefont {Vishwanath}},\
  }\href {\doibase 10.1103/PhysRevX.11.031005} {\bibfield  {journal} {\bibinfo
  {journal} {Phys. Rev. X}\ }\textbf {\bibinfo {volume} {11}},\ \bibinfo
  {pages} {031005} (\bibinfo {year} {2021})}\BibitemShut {NoStop}%
\bibitem [{\citenamefont {Satzinger}\ \emph {et~al.}(2021)\citenamefont
  {Satzinger} \emph {et~al.}}]{Google}%
  \BibitemOpen
  \bibfield  {author} {\bibinfo {author} {\bibfnamefont {K.~J.}\ \bibnamefont
  {Satzinger}} \emph {et~al.},\ }\href {\doibase 10.1126/science.abi8378}
  {\bibfield  {journal} {\bibinfo  {journal} {Science}\ }\textbf {\bibinfo
  {volume} {374}},\ \bibinfo {pages} {1237} (\bibinfo {year}
  {2021})}\BibitemShut {NoStop}%
\bibitem [{\citenamefont {Semeghini}\ \emph {et~al.}(2021)\citenamefont
  {Semeghini}, \citenamefont {Levine}, \citenamefont {Keesling}, \citenamefont
  {Ebadi}, \citenamefont {Wang}, \citenamefont {Bluvstein}, \citenamefont
  {Verresen}, \citenamefont {Pichler}, \citenamefont {Kalinowski},
  \citenamefont {Samajdar}, \citenamefont {Omran}, \citenamefont {Sachdev},
  \citenamefont {Vishwanath}, \citenamefont {Greiner}, \citenamefont
  {Vuletić},\ and\ \citenamefont {Lukin}}]{Lukin}%
  \BibitemOpen
  \bibfield  {author} {\bibinfo {author} {\bibfnamefont {G.}~\bibnamefont
  {Semeghini}}, \bibinfo {author} {\bibfnamefont {H.}~\bibnamefont {Levine}},
  \bibinfo {author} {\bibfnamefont {A.}~\bibnamefont {Keesling}}, \bibinfo
  {author} {\bibfnamefont {S.}~\bibnamefont {Ebadi}}, \bibinfo {author}
  {\bibfnamefont {T.~T.}\ \bibnamefont {Wang}}, \bibinfo {author}
  {\bibfnamefont {D.}~\bibnamefont {Bluvstein}}, \bibinfo {author}
  {\bibfnamefont {R.}~\bibnamefont {Verresen}}, \bibinfo {author}
  {\bibfnamefont {H.}~\bibnamefont {Pichler}}, \bibinfo {author} {\bibfnamefont
  {M.}~\bibnamefont {Kalinowski}}, \bibinfo {author} {\bibfnamefont
  {R.}~\bibnamefont {Samajdar}}, \bibinfo {author} {\bibfnamefont
  {A.}~\bibnamefont {Omran}}, \bibinfo {author} {\bibfnamefont
  {S.}~\bibnamefont {Sachdev}}, \bibinfo {author} {\bibfnamefont
  {A.}~\bibnamefont {Vishwanath}}, \bibinfo {author} {\bibfnamefont
  {M.}~\bibnamefont {Greiner}}, \bibinfo {author} {\bibfnamefont
  {V.}~\bibnamefont {Vuletić}}, \ and\ \bibinfo {author} {\bibfnamefont
  {M.~D.}\ \bibnamefont {Lukin}},\ }\href {\doibase 10.1126/science.abi8794}
  {\bibfield  {journal} {\bibinfo  {journal} {Science}\ }\textbf {\bibinfo
  {volume} {374}},\ \bibinfo {pages} {1242–1247} (\bibinfo {year}
  {2021})}\BibitemShut {NoStop}%
\bibitem [{\citenamefont {Stormer}\ \emph {et~al.}(1999)\citenamefont
  {Stormer}, \citenamefont {Tsui},\ and\ \citenamefont
  {Gossard}}]{stormer1999fractional}%
  \BibitemOpen
  \bibfield  {author} {\bibinfo {author} {\bibfnamefont {H.~L.}\ \bibnamefont
  {Stormer}}, \bibinfo {author} {\bibfnamefont {D.~C.}\ \bibnamefont {Tsui}}, \
  and\ \bibinfo {author} {\bibfnamefont {A.~C.}\ \bibnamefont {Gossard}},\
  }\href@noop {} {\bibfield  {journal} {\bibinfo  {journal} {Reviews of Modern
  Physics}\ }\textbf {\bibinfo {volume} {71}},\ \bibinfo {pages} {S298}
  (\bibinfo {year} {1999})}\BibitemShut {NoStop}%
\bibitem [{\citenamefont {Sarma}\ and\ \citenamefont {Pinczuk}(2008)}]{Girvin}%
  \BibitemOpen
  \bibfield  {author} {\bibinfo {author} {\bibfnamefont {S.~D.}\ \bibnamefont
  {Sarma}}\ and\ \bibinfo {author} {\bibfnamefont {A.}~\bibnamefont
  {Pinczuk}},\ }\href@noop {} {\emph {\bibinfo {title} {Perspectives in quantum
  hall effects: Novel quantum liquids in low-dimensional semiconductor
  structures}}}\ (\bibinfo  {publisher} {John Wiley \& Sons},\ \bibinfo {year}
  {2008})\BibitemShut {NoStop}%
\bibitem [{\citenamefont {Halperin}(1983)}]{Halperin1983}%
  \BibitemOpen
  \bibfield  {author} {\bibinfo {author} {\bibfnamefont {B.~I.}\ \bibnamefont
  {Halperin}},\ }\href@noop {} {\bibfield  {journal} {\bibinfo  {journal}
  {Helvetica Physica Acta}\ }\textbf {\bibinfo {volume} {56}},\ \bibinfo
  {pages} {75} (\bibinfo {year} {1983})}\BibitemShut {NoStop}%
\bibitem [{\citenamefont {Eisenstein}\ and\ \citenamefont
  {MacDonald}(2004)}]{Eisenstein2004}%
  \BibitemOpen
  \bibfield  {author} {\bibinfo {author} {\bibfnamefont {J.}~\bibnamefont
  {Eisenstein}}\ and\ \bibinfo {author} {\bibfnamefont {A.}~\bibnamefont
  {MacDonald}},\ }\href@noop {} {\bibfield  {journal} {\bibinfo  {journal}
  {Nature}\ }\textbf {\bibinfo {volume} {432}},\ \bibinfo {pages} {691}
  (\bibinfo {year} {2004})}\BibitemShut {NoStop}%
\bibitem [{\citenamefont {Eisenstein}(2014{\natexlab{a}})}]{Eisenstein2014}%
  \BibitemOpen
  \bibfield  {author} {\bibinfo {author} {\bibfnamefont {J.}~\bibnamefont
  {Eisenstein}},\ }\href@noop {} {\bibfield  {journal} {\bibinfo  {journal}
  {Annu. Rev. Condens. Matter Phys.}\ }\textbf {\bibinfo {volume} {5}},\
  \bibinfo {pages} {159} (\bibinfo {year} {2014}{\natexlab{a}})}\BibitemShut
  {NoStop}%
\bibitem [{\citenamefont {Moon}\ \emph {et~al.}(1995)\citenamefont {Moon},
  \citenamefont {Mori}, \citenamefont {Yang}, \citenamefont {Girvin},
  \citenamefont {MacDonald}, \citenamefont {Zheng}, \citenamefont {Yoshioka},\
  and\ \citenamefont {Zhang}}]{moon1995spontaneous}%
  \BibitemOpen
  \bibfield  {author} {\bibinfo {author} {\bibfnamefont {K.}~\bibnamefont
  {Moon}}, \bibinfo {author} {\bibfnamefont {H.}~\bibnamefont {Mori}}, \bibinfo
  {author} {\bibfnamefont {K.}~\bibnamefont {Yang}}, \bibinfo {author}
  {\bibfnamefont {S.}~\bibnamefont {Girvin}}, \bibinfo {author} {\bibfnamefont
  {A.}~\bibnamefont {MacDonald}}, \bibinfo {author} {\bibfnamefont
  {L.}~\bibnamefont {Zheng}}, \bibinfo {author} {\bibfnamefont
  {D.}~\bibnamefont {Yoshioka}}, \ and\ \bibinfo {author} {\bibfnamefont
  {S.-C.}\ \bibnamefont {Zhang}},\ }\href@noop {} {\bibfield  {journal}
  {\bibinfo  {journal} {Physical Review B}\ }\textbf {\bibinfo {volume} {51}},\
  \bibinfo {pages} {5138} (\bibinfo {year} {1995})}\BibitemShut {NoStop}%
\bibitem [{\citenamefont {Yang}\ \emph {et~al.}(1996)\citenamefont {Yang},
  \citenamefont {Moon}, \citenamefont {Belkhir}, \citenamefont {Mori},
  \citenamefont {Girvin}, \citenamefont {MacDonald}, \citenamefont {Zheng},\
  and\ \citenamefont {Yoshioka}}]{yang1996spontaneous}%
  \BibitemOpen
  \bibfield  {author} {\bibinfo {author} {\bibfnamefont {K.}~\bibnamefont
  {Yang}}, \bibinfo {author} {\bibfnamefont {K.}~\bibnamefont {Moon}}, \bibinfo
  {author} {\bibfnamefont {L.}~\bibnamefont {Belkhir}}, \bibinfo {author}
  {\bibfnamefont {H.}~\bibnamefont {Mori}}, \bibinfo {author} {\bibfnamefont
  {S.}~\bibnamefont {Girvin}}, \bibinfo {author} {\bibfnamefont
  {A.}~\bibnamefont {MacDonald}}, \bibinfo {author} {\bibfnamefont
  {L.}~\bibnamefont {Zheng}}, \ and\ \bibinfo {author} {\bibfnamefont
  {D.}~\bibnamefont {Yoshioka}},\ }\href@noop {} {\bibfield  {journal}
  {\bibinfo  {journal} {Physical Review B}\ }\textbf {\bibinfo {volume} {54}},\
  \bibinfo {pages} {11644} (\bibinfo {year} {1996})}\BibitemShut {NoStop}%
\bibitem [{\citenamefont {Liu}\ \emph {et~al.}(2017)\citenamefont {Liu},
  \citenamefont {Watanabe}, \citenamefont {Taniguchi}, \citenamefont
  {Halperin},\ and\ \citenamefont {Kim}}]{liu2017quantum}%
  \BibitemOpen
  \bibfield  {author} {\bibinfo {author} {\bibfnamefont {X.}~\bibnamefont
  {Liu}}, \bibinfo {author} {\bibfnamefont {K.}~\bibnamefont {Watanabe}},
  \bibinfo {author} {\bibfnamefont {T.}~\bibnamefont {Taniguchi}}, \bibinfo
  {author} {\bibfnamefont {B.~I.}\ \bibnamefont {Halperin}}, \ and\ \bibinfo
  {author} {\bibfnamefont {P.}~\bibnamefont {Kim}},\ }\href@noop {} {\bibfield
  {journal} {\bibinfo  {journal} {Nature Physics}\ }\textbf {\bibinfo {volume}
  {13}},\ \bibinfo {pages} {746} (\bibinfo {year} {2017})}\BibitemShut
  {NoStop}%
\bibitem [{\citenamefont {Li}\ \emph {et~al.}(2017)\citenamefont {Li},
  \citenamefont {Taniguchi}, \citenamefont {Watanabe}, \citenamefont {Hone},\
  and\ \citenamefont {Dean}}]{li2017excitonic}%
  \BibitemOpen
  \bibfield  {author} {\bibinfo {author} {\bibfnamefont {J.}~\bibnamefont
  {Li}}, \bibinfo {author} {\bibfnamefont {T.}~\bibnamefont {Taniguchi}},
  \bibinfo {author} {\bibfnamefont {K.}~\bibnamefont {Watanabe}}, \bibinfo
  {author} {\bibfnamefont {J.}~\bibnamefont {Hone}}, \ and\ \bibinfo {author}
  {\bibfnamefont {C.}~\bibnamefont {Dean}},\ }\href@noop {} {\bibfield
  {journal} {\bibinfo  {journal} {Nature Physics}\ }\textbf {\bibinfo {volume}
  {13}},\ \bibinfo {pages} {751} (\bibinfo {year} {2017})}\BibitemShut
  {NoStop}%
\bibitem [{\citenamefont {Bonesteel}\ \emph {et~al.}(1996)\citenamefont
  {Bonesteel}, \citenamefont {McDonald},\ and\ \citenamefont
  {Nayak}}]{bonesteel1996gauge}%
  \BibitemOpen
  \bibfield  {author} {\bibinfo {author} {\bibfnamefont {N.}~\bibnamefont
  {Bonesteel}}, \bibinfo {author} {\bibfnamefont {I.}~\bibnamefont {McDonald}},
  \ and\ \bibinfo {author} {\bibfnamefont {C.}~\bibnamefont {Nayak}},\
  }\href@noop {} {\bibfield  {journal} {\bibinfo  {journal} {Physical review
  letters}\ }\textbf {\bibinfo {volume} {77}},\ \bibinfo {pages} {3009}
  (\bibinfo {year} {1996})}\BibitemShut {NoStop}%
\bibitem [{\citenamefont {Kim}\ \emph {et~al.}(2001)\citenamefont {Kim},
  \citenamefont {Nayak}, \citenamefont {Demler}, \citenamefont {Read},\ and\
  \citenamefont {Sarma}}]{kim2001bilayer}%
  \BibitemOpen
  \bibfield  {author} {\bibinfo {author} {\bibfnamefont {Y.~B.}\ \bibnamefont
  {Kim}}, \bibinfo {author} {\bibfnamefont {C.}~\bibnamefont {Nayak}}, \bibinfo
  {author} {\bibfnamefont {E.}~\bibnamefont {Demler}}, \bibinfo {author}
  {\bibfnamefont {N.}~\bibnamefont {Read}}, \ and\ \bibinfo {author}
  {\bibfnamefont {S.~D.}\ \bibnamefont {Sarma}},\ }\href@noop {} {\bibfield
  {journal} {\bibinfo  {journal} {Physical Review B}\ }\textbf {\bibinfo
  {volume} {63}},\ \bibinfo {pages} {205315} (\bibinfo {year}
  {2001})}\BibitemShut {NoStop}%
\bibitem [{\citenamefont {Schliemann}\ \emph {et~al.}(2001)\citenamefont
  {Schliemann}, \citenamefont {Girvin},\ and\ \citenamefont
  {MacDonald}}]{schliemann2001strong}%
  \BibitemOpen
  \bibfield  {author} {\bibinfo {author} {\bibfnamefont {J.}~\bibnamefont
  {Schliemann}}, \bibinfo {author} {\bibfnamefont {S.}~\bibnamefont {Girvin}},
  \ and\ \bibinfo {author} {\bibfnamefont {A.}~\bibnamefont {MacDonald}},\
  }\href@noop {} {\bibfield  {journal} {\bibinfo  {journal} {Physical Review
  Letters}\ }\textbf {\bibinfo {volume} {86}},\ \bibinfo {pages} {1849}
  (\bibinfo {year} {2001})}\BibitemShut {NoStop}%
\bibitem [{\citenamefont {Stern}\ and\ \citenamefont
  {Halperin}(2002)}]{stern2002strong}%
  \BibitemOpen
  \bibfield  {author} {\bibinfo {author} {\bibfnamefont {A.}~\bibnamefont
  {Stern}}\ and\ \bibinfo {author} {\bibfnamefont {B.}~\bibnamefont
  {Halperin}},\ }\href@noop {} {\bibfield  {journal} {\bibinfo  {journal}
  {Physical review letters}\ }\textbf {\bibinfo {volume} {88}},\ \bibinfo
  {pages} {106801} (\bibinfo {year} {2002})}\BibitemShut {NoStop}%
\bibitem [{\citenamefont {Simon}\ \emph {et~al.}(2003)\citenamefont {Simon},
  \citenamefont {Rezayi},\ and\ \citenamefont
  {Milovanovic}}]{simon2003coexistence}%
  \BibitemOpen
  \bibfield  {author} {\bibinfo {author} {\bibfnamefont {S.~H.}\ \bibnamefont
  {Simon}}, \bibinfo {author} {\bibfnamefont {E.}~\bibnamefont {Rezayi}}, \
  and\ \bibinfo {author} {\bibfnamefont {M.~V.}\ \bibnamefont {Milovanovic}},\
  }\href@noop {} {\bibfield  {journal} {\bibinfo  {journal} {Physical review
  letters}\ }\textbf {\bibinfo {volume} {91}},\ \bibinfo {pages} {046803}
  (\bibinfo {year} {2003})}\BibitemShut {NoStop}%
\bibitem [{\citenamefont {Sheng}\ \emph {et~al.}(2003)\citenamefont {Sheng},
  \citenamefont {Balents},\ and\ \citenamefont {Wang}}]{sheng2003phase}%
  \BibitemOpen
  \bibfield  {author} {\bibinfo {author} {\bibfnamefont {D.}~\bibnamefont
  {Sheng}}, \bibinfo {author} {\bibfnamefont {L.}~\bibnamefont {Balents}}, \
  and\ \bibinfo {author} {\bibfnamefont {Z.}~\bibnamefont {Wang}},\ }\href@noop
  {} {\bibfield  {journal} {\bibinfo  {journal} {Physical review letters}\
  }\textbf {\bibinfo {volume} {91}},\ \bibinfo {pages} {116802} (\bibinfo
  {year} {2003})}\BibitemShut {NoStop}%
\bibitem [{\citenamefont {Park}(2004)}]{park2004spontaneous}%
  \BibitemOpen
  \bibfield  {author} {\bibinfo {author} {\bibfnamefont {K.}~\bibnamefont
  {Park}},\ }\href@noop {} {\bibfield  {journal} {\bibinfo  {journal} {Physical
  Review B}\ }\textbf {\bibinfo {volume} {69}},\ \bibinfo {pages} {045319}
  (\bibinfo {year} {2004})}\BibitemShut {NoStop}%
\bibitem [{\citenamefont {Shibata}\ and\ \citenamefont
  {Yoshioka}(2006)}]{shibata2006ground}%
  \BibitemOpen
  \bibfield  {author} {\bibinfo {author} {\bibfnamefont {N.}~\bibnamefont
  {Shibata}}\ and\ \bibinfo {author} {\bibfnamefont {D.}~\bibnamefont
  {Yoshioka}},\ }\href@noop {} {\bibfield  {journal} {\bibinfo  {journal}
  {Journal of the Physical Society of Japan}\ }\textbf {\bibinfo {volume}
  {75}},\ \bibinfo {pages} {043712} (\bibinfo {year} {2006})}\BibitemShut
  {NoStop}%
\bibitem [{\citenamefont {M{\"o}ller}\ \emph {et~al.}(2008)\citenamefont
  {M{\"o}ller}, \citenamefont {Simon},\ and\ \citenamefont
  {Rezayi}}]{moller2008paired}%
  \BibitemOpen
  \bibfield  {author} {\bibinfo {author} {\bibfnamefont {G.}~\bibnamefont
  {M{\"o}ller}}, \bibinfo {author} {\bibfnamefont {S.~H.}\ \bibnamefont
  {Simon}}, \ and\ \bibinfo {author} {\bibfnamefont {E.~H.}\ \bibnamefont
  {Rezayi}},\ }\href@noop {} {\bibfield  {journal} {\bibinfo  {journal}
  {Physical review letters}\ }\textbf {\bibinfo {volume} {101}},\ \bibinfo
  {pages} {176803} (\bibinfo {year} {2008})}\BibitemShut {NoStop}%
\bibitem [{\citenamefont {M{\"o}ller}\ \emph {et~al.}(2009)\citenamefont
  {M{\"o}ller}, \citenamefont {Simon},\ and\ \citenamefont
  {Rezayi}}]{moller2009trial}%
  \BibitemOpen
  \bibfield  {author} {\bibinfo {author} {\bibfnamefont {G.}~\bibnamefont
  {M{\"o}ller}}, \bibinfo {author} {\bibfnamefont {S.~H.}\ \bibnamefont
  {Simon}}, \ and\ \bibinfo {author} {\bibfnamefont {E.~H.}\ \bibnamefont
  {Rezayi}},\ }\href@noop {} {\bibfield  {journal} {\bibinfo  {journal}
  {Physical Review B}\ }\textbf {\bibinfo {volume} {79}},\ \bibinfo {pages}
  {125106} (\bibinfo {year} {2009})}\BibitemShut {NoStop}%
\bibitem [{\citenamefont {Milovanovi{\'c}}\ and\ \citenamefont
  {Papi{\'c}}(2009)}]{milovanovic2009nonperturbative}%
  \BibitemOpen
  \bibfield  {author} {\bibinfo {author} {\bibfnamefont {M.}~\bibnamefont
  {Milovanovi{\'c}}}\ and\ \bibinfo {author} {\bibfnamefont {Z.}~\bibnamefont
  {Papi{\'c}}},\ }\href@noop {} {\bibfield  {journal} {\bibinfo  {journal}
  {Physical Review B}\ }\textbf {\bibinfo {volume} {79}},\ \bibinfo {pages}
  {115319} (\bibinfo {year} {2009})}\BibitemShut {NoStop}%
\bibitem [{\citenamefont {Alicea}\ \emph {et~al.}(2009)\citenamefont {Alicea},
  \citenamefont {Motrunich}, \citenamefont {Refael},\ and\ \citenamefont
  {Fisher}}]{alicea2009interlayer}%
  \BibitemOpen
  \bibfield  {author} {\bibinfo {author} {\bibfnamefont {J.}~\bibnamefont
  {Alicea}}, \bibinfo {author} {\bibfnamefont {O.~I.}\ \bibnamefont
  {Motrunich}}, \bibinfo {author} {\bibfnamefont {G.}~\bibnamefont {Refael}}, \
  and\ \bibinfo {author} {\bibfnamefont {M.~P.}\ \bibnamefont {Fisher}},\
  }\href@noop {} {\bibfield  {journal} {\bibinfo  {journal} {Physical review
  letters}\ }\textbf {\bibinfo {volume} {103}},\ \bibinfo {pages} {256403}
  (\bibinfo {year} {2009})}\BibitemShut {NoStop}%
\bibitem [{\citenamefont {Papi{\'c}}\ and\ \citenamefont
  {Milovanovi{\'c}}(2012)}]{papic2012disordering}%
  \BibitemOpen
  \bibfield  {author} {\bibinfo {author} {\bibfnamefont {Z.}~\bibnamefont
  {Papi{\'c}}}\ and\ \bibinfo {author} {\bibfnamefont {M.}~\bibnamefont
  {Milovanovi{\'c}}},\ }\href@noop {} {\bibfield  {journal} {\bibinfo
  {journal} {Modern Physics Letters B}\ }\textbf {\bibinfo {volume} {26}},\
  \bibinfo {pages} {1250134} (\bibinfo {year} {2012})}\BibitemShut {NoStop}%
\bibitem [{\citenamefont {Sodemann}\ \emph {et~al.}(2017)\citenamefont
  {Sodemann}, \citenamefont {Kimchi}, \citenamefont {Wang},\ and\ \citenamefont
  {Senthil}}]{sodemann2017composite}%
  \BibitemOpen
  \bibfield  {author} {\bibinfo {author} {\bibfnamefont {I.}~\bibnamefont
  {Sodemann}}, \bibinfo {author} {\bibfnamefont {I.}~\bibnamefont {Kimchi}},
  \bibinfo {author} {\bibfnamefont {C.}~\bibnamefont {Wang}}, \ and\ \bibinfo
  {author} {\bibfnamefont {T.}~\bibnamefont {Senthil}},\ }\href@noop {}
  {\bibfield  {journal} {\bibinfo  {journal} {Physical Review B}\ }\textbf
  {\bibinfo {volume} {95}},\ \bibinfo {pages} {085135} (\bibinfo {year}
  {2017})}\BibitemShut {NoStop}%
\bibitem [{\citenamefont {Isobe}\ and\ \citenamefont
  {Fu}(2017)}]{isobe2017interlayer}%
  \BibitemOpen
  \bibfield  {author} {\bibinfo {author} {\bibfnamefont {H.}~\bibnamefont
  {Isobe}}\ and\ \bibinfo {author} {\bibfnamefont {L.}~\bibnamefont {Fu}},\
  }\href@noop {} {\bibfield  {journal} {\bibinfo  {journal} {Physical Review
  Letters}\ }\textbf {\bibinfo {volume} {118}},\ \bibinfo {pages} {166401}
  (\bibinfo {year} {2017})}\BibitemShut {NoStop}%
\bibitem [{\citenamefont {Zhu}\ \emph {et~al.}(2017)\citenamefont {Zhu},
  \citenamefont {Fu},\ and\ \citenamefont {Sheng}}]{zhu2017numerical}%
  \BibitemOpen
  \bibfield  {author} {\bibinfo {author} {\bibfnamefont {Z.}~\bibnamefont
  {Zhu}}, \bibinfo {author} {\bibfnamefont {L.}~\bibnamefont {Fu}}, \ and\
  \bibinfo {author} {\bibfnamefont {D.}~\bibnamefont {Sheng}},\ }\href@noop {}
  {\bibfield  {journal} {\bibinfo  {journal} {Physical review letters}\
  }\textbf {\bibinfo {volume} {119}},\ \bibinfo {pages} {177601} (\bibinfo
  {year} {2017})}\BibitemShut {NoStop}%
\bibitem [{\citenamefont {Lian}\ and\ \citenamefont
  {Zhang}(2018)}]{lian2018wave}%
  \BibitemOpen
  \bibfield  {author} {\bibinfo {author} {\bibfnamefont {B.}~\bibnamefont
  {Lian}}\ and\ \bibinfo {author} {\bibfnamefont {S.-C.}\ \bibnamefont
  {Zhang}},\ }\href@noop {} {\bibfield  {journal} {\bibinfo  {journal}
  {Physical review letters}\ }\textbf {\bibinfo {volume} {120}},\ \bibinfo
  {pages} {077601} (\bibinfo {year} {2018})}\BibitemShut {NoStop}%
\bibitem [{\citenamefont {Wagner}\ \emph {et~al.}(2021)\citenamefont {Wagner},
  \citenamefont {Nguyen}, \citenamefont {Simon},\ and\ \citenamefont
  {Halperin}}]{wagner2021s}%
  \BibitemOpen
  \bibfield  {author} {\bibinfo {author} {\bibfnamefont {G.}~\bibnamefont
  {Wagner}}, \bibinfo {author} {\bibfnamefont {D.~X.}\ \bibnamefont {Nguyen}},
  \bibinfo {author} {\bibfnamefont {S.~H.}\ \bibnamefont {Simon}}, \ and\
  \bibinfo {author} {\bibfnamefont {B.~I.}\ \bibnamefont {Halperin}},\
  }\href@noop {} {\bibfield  {journal} {\bibinfo  {journal} {Physical Review
  Letters}\ }\textbf {\bibinfo {volume} {127}},\ \bibinfo {pages} {246803}
  (\bibinfo {year} {2021})}\BibitemShut {NoStop}%
\bibitem [{\citenamefont {Liu}\ \emph {et~al.}(2022)\citenamefont {Liu},
  \citenamefont {Li}, \citenamefont {Watanabe}, \citenamefont {Taniguchi},
  \citenamefont {Hone}, \citenamefont {Halperin}, \citenamefont {Kim},\ and\
  \citenamefont {Dean}}]{liu2022crossover}%
  \BibitemOpen
  \bibfield  {author} {\bibinfo {author} {\bibfnamefont {X.}~\bibnamefont
  {Liu}}, \bibinfo {author} {\bibfnamefont {J.}~\bibnamefont {Li}}, \bibinfo
  {author} {\bibfnamefont {K.}~\bibnamefont {Watanabe}}, \bibinfo {author}
  {\bibfnamefont {T.}~\bibnamefont {Taniguchi}}, \bibinfo {author}
  {\bibfnamefont {J.}~\bibnamefont {Hone}}, \bibinfo {author} {\bibfnamefont
  {B.~I.}\ \bibnamefont {Halperin}}, \bibinfo {author} {\bibfnamefont
  {P.}~\bibnamefont {Kim}}, \ and\ \bibinfo {author} {\bibfnamefont {C.~R.}\
  \bibnamefont {Dean}},\ }\href@noop {} {\bibfield  {journal} {\bibinfo
  {journal} {Science}\ }\textbf {\bibinfo {volume} {375}},\ \bibinfo {pages}
  {205} (\bibinfo {year} {2022})}\BibitemShut {NoStop}%
\bibitem [{\citenamefont {Yang}(2001)}]{yang2001dipolar}%
  \BibitemOpen
  \bibfield  {author} {\bibinfo {author} {\bibfnamefont {K.}~\bibnamefont
  {Yang}},\ }\href@noop {} {\bibfield  {journal} {\bibinfo  {journal} {Physical
  Review Letters}\ }\textbf {\bibinfo {volume} {87}},\ \bibinfo {pages}
  {056802} (\bibinfo {year} {2001})}\BibitemShut {NoStop}%
\bibitem [{\citenamefont {Laughlin}(1983)}]{Laughlin83}%
  \BibitemOpen
  \bibfield  {author} {\bibinfo {author} {\bibfnamefont {R.~B.}\ \bibnamefont
  {Laughlin}},\ }\href@noop {} {\bibfield  {journal} {\bibinfo  {journal}
  {Physical Review Letters}\ }\textbf {\bibinfo {volume} {50}},\ \bibinfo
  {pages} {1395} (\bibinfo {year} {1983})}\BibitemShut {NoStop}%
\bibitem [{\citenamefont {Zeng}(2021)}]{zeng2021study}%
  \BibitemOpen
  \bibfield  {author} {\bibinfo {author} {\bibfnamefont {Y.}~\bibnamefont
  {Zeng}},\ }\emph {\bibinfo {title} {Study of Two-dimensional Correlated
  Quantum Fluid in Multi-layer graphene system}},\ \href@noop {} {Ph.D.
  thesis},\ \bibinfo  {school} {Columbia University} (\bibinfo {year}
  {2021})\BibitemShut {NoStop}%
\bibitem [{\citenamefont {Champagne}\ \emph {et~al.}(2008)\citenamefont
  {Champagne}, \citenamefont {Finck}, \citenamefont {Eisenstein}, \citenamefont
  {Pfeiffer},\ and\ \citenamefont {West}}]{champagne2008charge}%
  \BibitemOpen
  \bibfield  {author} {\bibinfo {author} {\bibfnamefont {A.}~\bibnamefont
  {Champagne}}, \bibinfo {author} {\bibfnamefont {A.}~\bibnamefont {Finck}},
  \bibinfo {author} {\bibfnamefont {J.}~\bibnamefont {Eisenstein}}, \bibinfo
  {author} {\bibfnamefont {L.}~\bibnamefont {Pfeiffer}}, \ and\ \bibinfo
  {author} {\bibfnamefont {K.}~\bibnamefont {West}},\ }\href@noop {} {\bibfield
   {journal} {\bibinfo  {journal} {Physical Review B}\ }\textbf {\bibinfo
  {volume} {78}},\ \bibinfo {pages} {205310} (\bibinfo {year}
  {2008})}\BibitemShut {NoStop}%
\bibitem [{\citenamefont {Chen}\ and\ \citenamefont
  {Yang}(2012)}]{chen2012interaction}%
  \BibitemOpen
  \bibfield  {author} {\bibinfo {author} {\bibfnamefont {H.}~\bibnamefont
  {Chen}}\ and\ \bibinfo {author} {\bibfnamefont {K.}~\bibnamefont {Yang}},\
  }\href@noop {} {\bibfield  {journal} {\bibinfo  {journal} {Physical Review
  B}\ }\textbf {\bibinfo {volume} {85}},\ \bibinfo {pages} {195113} (\bibinfo
  {year} {2012})}\BibitemShut {NoStop}%
\bibitem [{\citenamefont {Haldane}(1985)}]{Haldane1985}%
  \BibitemOpen
  \bibfield  {author} {\bibinfo {author} {\bibfnamefont {F.}~\bibnamefont
  {Haldane}},\ }\href@noop {} {\bibfield  {journal} {\bibinfo  {journal}
  {Physical review letters}\ }\textbf {\bibinfo {volume} {55}},\ \bibinfo
  {pages} {2095} (\bibinfo {year} {1985})}\BibitemShut {NoStop}%
\bibitem [{\citenamefont {Metlitski}(2022)}]{metlitski2022boundary}%
  \BibitemOpen
  \bibfield  {author} {\bibinfo {author} {\bibfnamefont {M.}~\bibnamefont
  {Metlitski}},\ }\href@noop {} {\bibfield  {journal} {\bibinfo  {journal}
  {SciPost Physics}\ }\textbf {\bibinfo {volume} {12}},\ \bibinfo {pages} {131}
  (\bibinfo {year} {2022})}\BibitemShut {NoStop}%
\bibitem [{\citenamefont {Zanardi}\ and\ \citenamefont
  {Paunkovi\ifmmode~\acute{c}\else \'{c}\fi{}}(2006)}]{Zanardi2006}%
  \BibitemOpen
  \bibfield  {author} {\bibinfo {author} {\bibfnamefont {P.}~\bibnamefont
  {Zanardi}}\ and\ \bibinfo {author} {\bibfnamefont {N.}~\bibnamefont
  {Paunkovi\ifmmode~\acute{c}\else \'{c}\fi{}}},\ }\href {\doibase
  10.1103/PhysRevE.74.031123} {\bibfield  {journal} {\bibinfo  {journal} {Phys.
  Rev. E}\ }\textbf {\bibinfo {volume} {74}},\ \bibinfo {pages} {031123}
  (\bibinfo {year} {2006})}\BibitemShut {NoStop}%
\bibitem [{\citenamefont {Gu}(2010)}]{Gu2010}%
  \BibitemOpen
  \bibfield  {author} {\bibinfo {author} {\bibfnamefont {S.-J.}\ \bibnamefont
  {Gu}},\ }\href@noop {} {\bibfield  {journal} {\bibinfo  {journal}
  {International Journal of Modern Physics B}\ }\textbf {\bibinfo {volume}
  {24}},\ \bibinfo {pages} {4371} (\bibinfo {year} {2010})}\BibitemShut
  {NoStop}%
\bibitem [{\citenamefont {Kivelson}\ \emph {et~al.}(1988)\citenamefont
  {Kivelson}, \citenamefont {Rokhsar},\ and\ \citenamefont
  {Sethna}}]{Kivelson}%
  \BibitemOpen
  \bibfield  {author} {\bibinfo {author} {\bibfnamefont {S.~A.}\ \bibnamefont
  {Kivelson}}, \bibinfo {author} {\bibfnamefont {D.~S.}\ \bibnamefont
  {Rokhsar}}, \ and\ \bibinfo {author} {\bibfnamefont {J.~P.}\ \bibnamefont
  {Sethna}},\ }\href {\doibase 10.1209/0295-5075/6/4/013} {\bibfield  {journal}
  {\bibinfo  {journal} {Europhysics Letters}\ }\textbf {\bibinfo {volume}
  {6}},\ \bibinfo {pages} {353} (\bibinfo {year} {1988})}\BibitemShut {NoStop}%
\bibitem [{\citenamefont {Hasenbusch}\ and\ \citenamefont
  {Vicari}(2011)}]{hasenbusch2011anisotropic}%
  \BibitemOpen
  \bibfield  {author} {\bibinfo {author} {\bibfnamefont {M.}~\bibnamefont
  {Hasenbusch}}\ and\ \bibinfo {author} {\bibfnamefont {E.}~\bibnamefont
  {Vicari}},\ }\href@noop {} {\bibfield  {journal} {\bibinfo  {journal}
  {Physical Review B}\ }\textbf {\bibinfo {volume} {84}},\ \bibinfo {pages}
  {125136} (\bibinfo {year} {2011})}\BibitemShut {NoStop}%
\bibitem [{\citenamefont {Wen}(1991)}]{wen1991edge}%
  \BibitemOpen
  \bibfield  {author} {\bibinfo {author} {\bibfnamefont {X.}~\bibnamefont
  {Wen}},\ }\href@noop {} {\bibfield  {journal} {\bibinfo  {journal} {Modern
  Physics Letters B}\ }\textbf {\bibinfo {volume} {5}},\ \bibinfo {pages} {39}
  (\bibinfo {year} {1991})}\BibitemShut {NoStop}%
\bibitem [{\citenamefont
  {Eisenstein}(2014{\natexlab{b}})}]{eisenstein2014exciton}%
  \BibitemOpen
  \bibfield  {author} {\bibinfo {author} {\bibfnamefont {J.}~\bibnamefont
  {Eisenstein}},\ }\href@noop {} {\bibfield  {journal} {\bibinfo  {journal}
  {Annu. Rev. Condens. Matter Phys.}\ }\textbf {\bibinfo {volume} {5}},\
  \bibinfo {pages} {159} (\bibinfo {year} {2014}{\natexlab{b}})}\BibitemShut
  {NoStop}%
\bibitem [{\citenamefont {Spielman}\ \emph {et~al.}(2001)\citenamefont
  {Spielman}, \citenamefont {Eisenstein}, \citenamefont {Pfeiffer},\ and\
  \citenamefont {West}}]{spielman2001observation}%
  \BibitemOpen
  \bibfield  {author} {\bibinfo {author} {\bibfnamefont {I.}~\bibnamefont
  {Spielman}}, \bibinfo {author} {\bibfnamefont {J.}~\bibnamefont
  {Eisenstein}}, \bibinfo {author} {\bibfnamefont {L.}~\bibnamefont
  {Pfeiffer}}, \ and\ \bibinfo {author} {\bibfnamefont {K.}~\bibnamefont
  {West}},\ }\href@noop {} {\bibfield  {journal} {\bibinfo  {journal} {Physical
  Review Letters}\ }\textbf {\bibinfo {volume} {87}},\ \bibinfo {pages}
  {036803} (\bibinfo {year} {2001})}\BibitemShut {NoStop}%
\bibitem [{\citenamefont {Stern}\ \emph {et~al.}(2001)\citenamefont {Stern},
  \citenamefont {Girvin}, \citenamefont {MacDonald},\ and\ \citenamefont
  {Ma}}]{stern2001theory}%
  \BibitemOpen
  \bibfield  {author} {\bibinfo {author} {\bibfnamefont {A.}~\bibnamefont
  {Stern}}, \bibinfo {author} {\bibfnamefont {S.~M.}\ \bibnamefont {Girvin}},
  \bibinfo {author} {\bibfnamefont {A.~H.}\ \bibnamefont {MacDonald}}, \ and\
  \bibinfo {author} {\bibfnamefont {N.}~\bibnamefont {Ma}},\ }\href@noop {}
  {\bibfield  {journal} {\bibinfo  {journal} {Physical Review Letters}\
  }\textbf {\bibinfo {volume} {86}},\ \bibinfo {pages} {1829} (\bibinfo {year}
  {2001})}\BibitemShut {NoStop}%
\bibitem [{Note1()}]{Note1}%
  \BibitemOpen
  \bibinfo {note} {In this case, it is important to align the two graphene
  layers so the inter-layer tunneling probes the spectral weight of the exciton
  order parameter at momentum $\protect \mathbf q=0$ and energy $\omega
  =eV$.}\BibitemShut {Stop}%
\bibitem [{\citenamefont {Hebard}\ and\ \citenamefont
  {Paalanen}(1990)}]{hebard1990magnetic}%
  \BibitemOpen
  \bibfield  {author} {\bibinfo {author} {\bibfnamefont {A.}~\bibnamefont
  {Hebard}}\ and\ \bibinfo {author} {\bibfnamefont {M.}~\bibnamefont
  {Paalanen}},\ }\href@noop {} {\bibfield  {journal} {\bibinfo  {journal}
  {Physical review letters}\ }\textbf {\bibinfo {volume} {65}},\ \bibinfo
  {pages} {927} (\bibinfo {year} {1990})}\BibitemShut {NoStop}%
\bibitem [{\citenamefont {Yazdani}\ and\ \citenamefont
  {Kapitulnik}(1995)}]{yazdani1995superconducting}%
  \BibitemOpen
  \bibfield  {author} {\bibinfo {author} {\bibfnamefont {A.}~\bibnamefont
  {Yazdani}}\ and\ \bibinfo {author} {\bibfnamefont {A.}~\bibnamefont
  {Kapitulnik}},\ }\href@noop {} {\bibfield  {journal} {\bibinfo  {journal}
  {Physical review letters}\ }\textbf {\bibinfo {volume} {74}},\ \bibinfo
  {pages} {3037} (\bibinfo {year} {1995})}\BibitemShut {NoStop}%
\bibitem [{\citenamefont {Markovi{\'c}}\ \emph {et~al.}(1998)\citenamefont
  {Markovi{\'c}}, \citenamefont {Christiansen},\ and\ \citenamefont
  {Goldman}}]{markovic1998thickness}%
  \BibitemOpen
  \bibfield  {author} {\bibinfo {author} {\bibfnamefont {N.}~\bibnamefont
  {Markovi{\'c}}}, \bibinfo {author} {\bibfnamefont {C.}~\bibnamefont
  {Christiansen}}, \ and\ \bibinfo {author} {\bibfnamefont {A.}~\bibnamefont
  {Goldman}},\ }\href@noop {} {\bibfield  {journal} {\bibinfo  {journal}
  {Physical review letters}\ }\textbf {\bibinfo {volume} {81}},\ \bibinfo
  {pages} {5217} (\bibinfo {year} {1998})}\BibitemShut {NoStop}%
\bibitem [{\citenamefont {Bielejec}\ and\ \citenamefont
  {Wu}(2002)}]{bielejec2002field}%
  \BibitemOpen
  \bibfield  {author} {\bibinfo {author} {\bibfnamefont {E.}~\bibnamefont
  {Bielejec}}\ and\ \bibinfo {author} {\bibfnamefont {W.}~\bibnamefont {Wu}},\
  }\href@noop {} {\bibfield  {journal} {\bibinfo  {journal} {Physical review
  letters}\ }\textbf {\bibinfo {volume} {88}},\ \bibinfo {pages} {206802}
  (\bibinfo {year} {2002})}\BibitemShut {NoStop}%
\bibitem [{\citenamefont {Hen}\ \emph {et~al.}(2021)\citenamefont {Hen},
  \citenamefont {Zhang}, \citenamefont {Shelukhin}, \citenamefont
  {Kapitulnik},\ and\ \citenamefont {Palevski}}]{hen2021superconductor}%
  \BibitemOpen
  \bibfield  {author} {\bibinfo {author} {\bibfnamefont {B.}~\bibnamefont
  {Hen}}, \bibinfo {author} {\bibfnamefont {X.}~\bibnamefont {Zhang}}, \bibinfo
  {author} {\bibfnamefont {V.}~\bibnamefont {Shelukhin}}, \bibinfo {author}
  {\bibfnamefont {A.}~\bibnamefont {Kapitulnik}}, \ and\ \bibinfo {author}
  {\bibfnamefont {A.}~\bibnamefont {Palevski}},\ }\href@noop {} {\bibfield
  {journal} {\bibinfo  {journal} {Proceedings of the National Academy of
  Sciences}\ }\textbf {\bibinfo {volume} {118}},\ \bibinfo {pages}
  {e2015970118} (\bibinfo {year} {2021})}\BibitemShut {NoStop}%
\bibitem [{\citenamefont {Giamarchi}(2003)}]{giamarchi2003quantum}%
  \BibitemOpen
  \bibfield  {author} {\bibinfo {author} {\bibfnamefont {T.}~\bibnamefont
  {Giamarchi}},\ }\href@noop {} {\emph {\bibinfo {title} {Quantum physics in
  one dimension}}},\ Vol.\ \bibinfo {volume} {121}\ (\bibinfo  {publisher}
  {Clarendon press},\ \bibinfo {year} {2003})\BibitemShut {NoStop}%
\bibitem [{\citenamefont {Kane}\ \emph {et~al.}(1994)\citenamefont {Kane},
  \citenamefont {Fisher},\ and\ \citenamefont
  {Polchinski}}]{kane1994randomness}%
  \BibitemOpen
  \bibfield  {author} {\bibinfo {author} {\bibfnamefont {C.}~\bibnamefont
  {Kane}}, \bibinfo {author} {\bibfnamefont {M.~P.}\ \bibnamefont {Fisher}}, \
  and\ \bibinfo {author} {\bibfnamefont {J.}~\bibnamefont {Polchinski}},\
  }\href@noop {} {\bibfield  {journal} {\bibinfo  {journal} {Physical review
  letters}\ }\textbf {\bibinfo {volume} {72}},\ \bibinfo {pages} {4129}
  (\bibinfo {year} {1994})}\BibitemShut {NoStop}%
\bibitem [{\citenamefont {Lou}\ \emph {et~al.}(2007)\citenamefont {Lou},
  \citenamefont {Sandvik},\ and\ \citenamefont {Balents}}]{Sandvik}%
  \BibitemOpen
  \bibfield  {author} {\bibinfo {author} {\bibfnamefont {J.}~\bibnamefont
  {Lou}}, \bibinfo {author} {\bibfnamefont {A.~W.}\ \bibnamefont {Sandvik}}, \
  and\ \bibinfo {author} {\bibfnamefont {L.}~\bibnamefont {Balents}},\ }\href
  {\doibase 10.1103/PhysRevLett.99.207203} {\bibfield  {journal} {\bibinfo
  {journal} {Phys. Rev. Lett.}\ }\textbf {\bibinfo {volume} {99}},\ \bibinfo
  {pages} {207203} (\bibinfo {year} {2007})}\BibitemShut {NoStop}%
\bibitem [{\citenamefont {Senthil}\ and\ \citenamefont
  {Fisher}(2001)}]{SenthilFisher}%
  \BibitemOpen
  \bibfield  {author} {\bibinfo {author} {\bibfnamefont {T.}~\bibnamefont
  {Senthil}}\ and\ \bibinfo {author} {\bibfnamefont {M.~P.~A.}\ \bibnamefont
  {Fisher}},\ }\href {\doibase 10.1103/physrevlett.86.292} {\bibfield
  {journal} {\bibinfo  {journal} {Physical Review Letters}\ }\textbf {\bibinfo
  {volume} {86}},\ \bibinfo {pages} {292} (\bibinfo {year} {2001})}\BibitemShut
  {NoStop}%
\bibitem [{\citenamefont {Wu}\ \emph {et~al.}(2023)\citenamefont {Wu},
  \citenamefont {Tu},\ and\ \citenamefont {Cheng}}]{wu2023continuous}%
  \BibitemOpen
  \bibfield  {author} {\bibinfo {author} {\bibfnamefont {Y.-H.}\ \bibnamefont
  {Wu}}, \bibinfo {author} {\bibfnamefont {H.-H.}\ \bibnamefont {Tu}}, \ and\
  \bibinfo {author} {\bibfnamefont {M.}~\bibnamefont {Cheng}},\ }\href@noop {}
  {\bibfield  {journal} {\bibinfo  {journal} {arXiv preprint arXiv:2302.06501}\
  } (\bibinfo {year} {2023})}\BibitemShut {NoStop}%
\bibitem [{\citenamefont {Levin}\ and\ \citenamefont {Stern}(2009)}]{Levin}%
  \BibitemOpen
  \bibfield  {author} {\bibinfo {author} {\bibfnamefont {M.}~\bibnamefont
  {Levin}}\ and\ \bibinfo {author} {\bibfnamefont {A.}~\bibnamefont {Stern}},\
  }\href {\doibase 10.1103/physrevlett.103.196803} {\bibfield  {journal}
  {\bibinfo  {journal} {Physical Review Letters}\ }\textbf {\bibinfo {volume}
  {103}} (\bibinfo {year} {2009}),\ 10.1103/physrevlett.103.196803}\BibitemShut
  {NoStop}%
\bibitem [{\citenamefont {Ghaemi}\ \emph {et~al.}(2012)\citenamefont {Ghaemi},
  \citenamefont {Cayssol}, \citenamefont {Sheng},\ and\ \citenamefont
  {Vishwanath}}]{Ghaemi}%
  \BibitemOpen
  \bibfield  {author} {\bibinfo {author} {\bibfnamefont {P.}~\bibnamefont
  {Ghaemi}}, \bibinfo {author} {\bibfnamefont {J.}~\bibnamefont {Cayssol}},
  \bibinfo {author} {\bibfnamefont {D.~N.}\ \bibnamefont {Sheng}}, \ and\
  \bibinfo {author} {\bibfnamefont {A.}~\bibnamefont {Vishwanath}},\ }\href
  {\doibase 10.1103/PhysRevLett.108.266801} {\bibfield  {journal} {\bibinfo
  {journal} {Phys. Rev. Lett.}\ }\textbf {\bibinfo {volume} {108}},\ \bibinfo
  {pages} {266801} (\bibinfo {year} {2012})}\BibitemShut {NoStop}%
\bibitem [{\citenamefont {Morf}\ and\ \citenamefont
  {Halperin}(1986)}]{HalperinMorf}%
  \BibitemOpen
  \bibfield  {author} {\bibinfo {author} {\bibfnamefont {R.}~\bibnamefont
  {Morf}}\ and\ \bibinfo {author} {\bibfnamefont {B.~I.}\ \bibnamefont
  {Halperin}},\ }\href {\doibase 10.1103/PhysRevB.33.2221} {\bibfield
  {journal} {\bibinfo  {journal} {Phys. Rev. B}\ }\textbf {\bibinfo {volume}
  {33}},\ \bibinfo {pages} {2221} (\bibinfo {year} {1986})}\BibitemShut
  {NoStop}%
\bibitem [{\citenamefont {Bonesteel}(1995)}]{Bonesteel}%
  \BibitemOpen
  \bibfield  {author} {\bibinfo {author} {\bibfnamefont {N.~E.}\ \bibnamefont
  {Bonesteel}},\ }\href {\doibase 10.1103/physrevb.51.9917} {\bibfield
  {journal} {\bibinfo  {journal} {Physical Review B}\ }\textbf {\bibinfo
  {volume} {51}},\ \bibinfo {pages} {9917} (\bibinfo {year}
  {1995})}\BibitemShut {NoStop}%
\end{thebibliography}%

\onecolumngrid

\appendix

\section{Numerical Details} 

We consider $\nu_T=1/3+2/3$ quantum Hall bilayers subject to a perpendicular magnetic field on torus,  which is spanned by length vectors $\mathbf{L_x}$ and $\mathbf{L_y}$, and thus the orbital number (or flux number) in each layer $N_\phi$ is determined by the area of torus, i.e., 
\begin{equation}\label{Eq:SNphi}
|\mathbf{L_x}\times \mathbf{L_y}|=2\pi N_\phi
\end{equation} 
Here,  the magnetic length $l_B\equiv\sqrt{\hbar c/eB}\equiv1$ (the unit of length). We choose Landau gauge $ \mathbf{A}=(By,0,0)$ and consider the torus with aspect ratio to be 1. The single-particle wave functions in the lowest Landau level (LLL) as basis reads
\begin{equation}
\psi_{k} (x,y) =\frac 1{\sqrt{L_x \sqrt{\pi}}}\sum_{n=-\infty}^{+\infty}e^{[\mathrm{i}(k+nL_y)x-(y+nL_y+k)^{2}/2]},
\label{wave}
\end{equation}
where $k\equiv  {2\pi j}/L_x $ with $j=0,1,...,N_\phi-1$ due to periodical boundary condition along $x$ direction. The single particle states $\psi_{k}$ are centered at $y=-k$ with a distance $2\pi/L_x$ apart along $y$ direction, while they are extended in $x$ direction. Then $N_\phi$ states can be mapped into one-dimensional (1D) lattice with each site representing a single particle orbital $\psi_{k}$. Then one can perform numerical simulation on such 1D lattice in momentum space with the number of sites equal to the number of orbitals. The relationship between the area of the torus and the size of 1D lattice is determined by Eq.\ref{Eq:SNphi}.
In order to realize the numerical diagonalization on a larger system size, one needs to reduce the dimension of the Hamiltonian block by taking advantage of magnetic translational symmetries along $x$ or/and $y$ directions. The symmetry analysis  was first provided by Haldane~\cite{Haldane1985} with introducing two translation operators,  $T_\alpha$ $(\alpha =1,2)$ with eigenvalues $e^{2\pi iK_{\lambda} /N_\phi}$ ($ \lambda=x,y$  and $ K_{\lambda}=0,...,N_\phi-1$).$T_1$ corresponds to the magnetic translation in $x$-direction, where $K_x=\sum_ {k=0}^{N_\phi-1} k n_k$ (mod $N_\phi$)  is total momentum (in the unit of $2\pi /L_x$) of electrons taken modulo $N_\phi$.  $T_2$ translates the entire lattice configuration one step $L_y/N_\phi=2\pi/L_x$ to the right along $y$-direction. Taking advantage of one or both symmetries, one can numerically diagonalize the Hamiltonian efficiently. Different from the sphere geometry, there is no orbital number shift on torus and the states are uniquely determined by their filling factor.  

In the present work, we  consider the physical systems with two identical 2D layers (with zero width) in the absence of  electron interlayer tunneling while  spins of electrons are fully polarized  due to strong magnetic fields. Such a system can be described by the projected Coulomb interaction,
 \begin{equation}\label {Ham}
V =\frac{1}{N_\phi} \sum\limits_{i < j,\alpha ,\beta } {\sum\limits_{{\bf{q}},{\bf{q}} \ne 0} {{V_{\alpha \beta }}\left( q \right)} } {e^{ -\frac{q^2}{2}}}L^2_n[\frac{q^2}{2}]{e^{i{\bf{q}} \cdot \left( {{{\bf{R}}_{\alpha ,i}} - {{\bf{R}}_{\beta ,j}}} \right)}}.
  \end{equation}
Here, $\alpha (\beta)=1,2$ denote two layers or, equivalently, two components of a pseudospin-$1/2$. $q$ =$|{\bf{q}}|$ =$ \sqrt {q_x^2 + q_y^2} $, $V_{11}(q)= V_{22}(q)={e^2}/({\varepsilon q})$ and $V_{12}(q)=V_{21}(q)= {e^2} /({\varepsilon q})\cdot e^{-qd}$  are the Fourier transformations of the intralayer and interlayer Coulomb interactions, respectively. $d$ represents the distance between two layers in the unit of magnetic length $l_B$. $L_n(x)$ is the Laguerre polynomial with Landau level index $n$ and $\bf{R}_{\alpha,i}$ is the  guiding center coordinate of the $i$-th electron in layer $\alpha$. Here we consider rectangular unit cells with $L_x=L_y=L$ and set magnetic length $l_B\equiv\sqrt{\hbar c/eB}$ as  the unit of length, $e^2/{\varepsilon l_B}$ as the unit of energy. Numerically, one needs to use the second-quantization form:
 \begin{equation}
 V = \sum\limits_{{j_1}{j_2}{j_3}{j_4}} {{V^{\alpha \beta }_{{j_1}{j_2}{j_3}{j_4}}}c_{{j_1}}^\dag c_{{j_2}}^\dag {c_{{j_3}}}{c_{{j_4}}}},
 \end{equation}
 with
 \begin{equation}\label{ED}
 {V^{\alpha \beta }_{{j_1}{j_2}{j_3}{j_4}}} = {{\delta '}_{{j_1} + {j_2},{j_3} + {j_4}}}\frac{1}{{4\pi  N_\phi}}\sum\limits_{{\bf{q}},{\bf{q}} \ne 0} {{{\delta '}_{{j_1} - {j_4},{{{q_y}{L_y}} \mathord{\left/
 {\vphantom {{{q_y}{L_y}} {2\pi }}} \right.
 \kern-\nulldelimiterspace} {2\pi }}}}} V_{\alpha \beta}(q)\exp \left[ {{{ - {q^2}} \mathord{\left/
 {\vphantom {{ - {q^2}} 2}} \right.
 \kern-\nulldelimiterspace} 2} - i\left( {{j_1} - {j_3}} \right){{{q_x}{L_x}} \mathord{\left/
 {\vphantom {{{q_x}{L_x}} {{N_\phi }}}} \right.
 \kern-\nulldelimiterspace} {{N_\phi }}}} \right] L^2_{n=0}[- {q^2}/2].
     \end{equation}
Here, the Kronecker delta with the prime means that the equation is defined modulo $N_\phi$.  We also consider a uniform and positive background charge so that the Coulomb interaction at $q=0$ is canceled out.

\section{Symmetry of the model}

Here we point out a symmetry $\mathcal M \mathcal C \mathcal T$ for  the quantum Hall bilayer at filling $(\nu_1,\nu_2)=(x,-x)$. Here $\nu_a=\frac{N_e}{N_{\Phi}}$, where $N_{\Phi}$ is the number of the magnetic flux in the system. $\nu_2<0$ means that the system is hole doped with hole density at $x$ per flux.   We will mainly be interested in the $x=1/3$ point.

We define electron operators in layer 1 and 2 as $c_1(\mathbf r)$ and $c_2(\mathbf r)$.  Because the layer 2 is hole doped, it is convenient to use the hole operator $h_2(\mathbf r)=c_2^\dagger(\mathbf r)$. The Hamiltonian is
\begin{equation}
\label{Eqn:ParticleHole2}
	\mathcal H= \int d^2 \mathbf r c^\dagger_1(\mathbf r)\frac{(-i\vec \nabla - e \vec A_1)^2}{2m} c_1(\mathbf r)+\int d^2 \mathbf r h^\dagger_2(\mathbf r)\frac{(-i\vec \nabla + e \vec A_2)^2}{2m}h_2(\mathbf r)-D \int d^2 \mathbf r \big( c^\dagger_1(\mathbf r)c_1(\mathbf r)+h^\dagger_2(\mathbf r) h_2(\mathbf r)\big)+\mathcal H_{\text{int}}
\end{equation}
where $e$ is  the electron charge which is negative.  $\vec A_1(\mathbf r)$ and $\vec A_2(\mathbf r)$ are vector fields in the two layers.  For quantum Hall bilayer system, we have the magnetic field: $\vec A_a(\mathbf r)=\frac{1}{2} B \hat z \times \mathbf r+\delta \vec A_a(\mathbf r)$.  Here $\delta \vec A_a(\mathbf r)$ is the probing field in each layer applied to measure the response of the system.  $D$ is the displacement field and can be viewed as the chemical potential to tune the exciton density $x$.

The interaction term is:

\begin{equation}
	\mathcal H_{\text{int}}=\frac{1}{2} \int d^2 \mathbf r \int d^2 \mathbf r' V_{ab}(|\mathbf r -\mathbf r'|) : \rho_a(\mathbf r) \rho_b(\mathbf r'):
\end{equation}
where
\begin{equation}
	\rho_1(\mathbf r)=e c^\dagger_1(\mathbf r) c_1(\mathbf r)
\end{equation}
and
\begin{equation}
	\rho_2(\mathbf r)=-e h^\dagger_2(\mathbf r)h_2(\mathbf r)
\end{equation}

The Coulomb interaction is in the form $V_{ab}(\mathbf r)=\frac{1}{(2\pi)^2}\int d^2 \mathbf q V_{ab}(\mathbf q)e^{i \mathbf q \cdot \mathbf r}$ with $V_{11}(q)= V_{22}(q)=\frac{e^2}{\varepsilon q}$ and $V_{12}(q)=V_{21}(q)= \frac{e^2}{\varepsilon q} e^{-qd}$. $d$ represents the distance between two layers in the unit of magnetic length  $l_B=\sqrt{\hbar c/eB}$.

Let us ignore the probing field $\delta \vec A_a(\mathbf r)$ for now.  Then the Hamiltonian has the following $\mathcal M  \mathcal C \mathcal T$ anti-unitary symmetry:

\begin{equation}
	\mathcal M \mathcal C \mathcal T:  c_1(\mathbf r) \leftrightarrow h_2(\mathbf r)
\end{equation}
Note that this is an anti-unitary and need to include a complex conjugate $\mathcal K$.  Under $\mathcal M \mathcal C \mathcal T$, $\rho_1(\mathbf r)\rightarrow -\rho_2(\mathbf r)$ and $\rho_1(\mathbf q)\rightarrow -\rho_2(-\mathbf q)$.   Under $\mathcal M \mathcal C \mathcal T$, an electron in layer 1 is transformed to a hole in layer 2. We can also include the vector potential $\delta \vec A_{a}(\mathbf r)$ and also the electric potential term $\delta H= -\rho_1(\mathbf r) A^0_1(\mathbf r)-\rho_2(\mathbf r)A^0_2(\mathbf r)$. Thus under $\mathcal M \mathcal C \mathcal T$, we have $A^0_1(\mathbf r) \rightarrow -A^0_2(\mathbf r)$ and $\delta \vec A_1(\mathbf r)\rightarrow \delta \vec A_2(\mathbf r)$.  In the following we use $\vec A_a(\mathbf r)$ to denote $\delta \vec A_a(\mathbf r)$.   The space time coordinates transform as $(t,\mathbf r)\rightarrow(-t,\mathbf r)$ under $\mathcal M \mathcal C \mathcal T$.

One can diagonalize the kinetic part and project the interaction in the lowest Landau levels.  Within the lowest Landau level, the Hamiltonian is

\begin{equation}
	\mathcal  H=\frac{1}{2} \sum_{a,b=1,2} V_{ab}(\mathbf q):  \rho_a(\mathbf q) \rho_a(-\mathbf q) :
\end{equation}

Here 
\begin{equation}
	\rho_a(\mathbf q)=\int d^2\mathbf q \rho_a(\mathbf r) e^{-i \mathbf q \cdot \mathbf r}
\end{equation}
with $\rho_a(\mathbf r)$ as the charge density operator projected to the lowest Landau level:
\begin{equation}
	\rho_1(\mathbf r)=e\sum_{m,n} c^\dagger_{1;m} c_{1;n}\varphi^*_{+;m}(\mathbf r) \varphi_{+;n}(\mathbf r)
\end{equation}
and
\begin{equation}
		\rho_2(\mathbf r)=-e\sum_{m,n} h^\dagger_{2;m} h_{2;n}\varphi^*_{-;m}(\mathbf r) \varphi_{-;n}(\mathbf r)
\end{equation}

In the above $\varphi_{+,m}(\mathbf r)$ and $\varphi_{-;m}$ are the wavefunction of  the state labeled by the Landau index $m$ for electron and hole respectively.  We have $\varphi_{-;m}(\mathbf r)=\varphi_{+;m}(\mathbf r)^*$.

Under the symmetry $\mathcal M  \mathcal C \mathcal T$, $c_{1;m} \rightarrow h_{2;m}$ and $\rho_1(\mathbf r) \rightarrow -\rho_2(\mathbf r)$ still hold in the Hamiltonian projected to the lowest Landau level.

\section{Relation to the FQSH to Paired-Superfluid Transition and an Estimate for $d_c$}
Thus far we have phrased the discussion in terms of exciton condensation on the $(\nu_1,\,\nu_2) = (1/3, \,2/3)$ insulator. Now consider performing a particle hole conjugation on Layer 2, as described above i.e. $h_2(r)=c^\dagger_2(r)$. This leads to the two layers now having the {\em same} density, but experiencing opposite magnetic fields as in Eqn. \ref{Eqn:ParticleHole2}. This is noting but the fractional version of the Quantum Spin Hall effect (FQSH) \cite{Levin}. Furthermore, as a result of the particle-hole transformation in the bottom layer, the repulsive inter-layer interaction now becomes an attractive interaction, and leads to the formation of bound states between the layers of Laughlin quasiparticles, i.e. Laughlin Cooper pairs of charge $2e/3$. The condensation of these fractional Cooper pairs leads to a paired superconductor \cite{Ghaemi} (which is smoothly connected to the paired condensate of electrons). Note, the Laughlin Cooper pair is a boson and has mutual statistics with all the other anyons in the problem, hence its condensation leads to a conventional superconductor. 

The following simple  energetic argument gives an estimate for the critical distance $d_c$. Consider  creating a Laughlin quasi-electron+quasi-hole in one layer and a corresponding pair in the other layer. The energy for each pair is just the gap: $\Delta_{1/3} \approx 0.1 e^2/\epsilon l_B$ \cite{HalperinMorf}. Now consider the strength of the attractive interaction between quasi-particle and quasi-hole in the two layers $\Delta E = {e^*}^2/\epsilon d$, where $e^*=e/3$. A simple estimate for the critical point is when the binding energy overcomes the cost of creating the quasiparticles and hence: ${e^*}^2/\epsilon d  = \Delta_{1/3}$ or $\frac{d_c^*}{l_B} =1.1 $. In practice the excitons will have a dispersion that further lowers their energy and we would expect that the true transition occurs earlier i.e. $d_c>d_c^*$. Our numerical calculations gave $d_c \approx 1.7$. Similarly one can estimate the critical distance for the $(\nu_1,\,\nu_2)=(1/5,\,4/5)$ using $\Delta_5=0.024e^2/l_B$ \cite{Bonesteel} and $e^*=e/5$.

\section{Critical theory}

We consider the filling $(\nu_1,\nu_2)=(\frac{1}{3},-\frac{1}{3})$. In the FQSH phase at large $d/l_B$, the effective low energy theory is captured by a Chern-Simons theory with K matrix $K=\begin{pmatrix}3 & 0 \\ 0 & -3\end{pmatrix}$. There are two emergent gauge fields $a_{1;\mu},a_{2;\mu}$ whose charges are labeled as $l=(l_1,l_2)$.  We consider a fractional exciton labeled by $l=(1,1)$. It carries physical charge $Q_c=0$ and physical spin $Q_s=\frac{1}{3}$. The transition between the FQSH phase and the exciton condensation phase is described by the condensation of this fractional exciton whose creation operator is labeled as $\varphi^\dagger$. Then the critical theory is:

\begin{align}
	\mathcal L&=|(\partial_\mu -i(a_{1;\mu}-a_{2;\mu}))\varphi|^2- s |\varphi|^2-g|\varphi|^4-\frac{3}{4\pi}a_1 d a_1+\frac{3}{4\pi}a_2 d a_2+\frac{1}{2\pi}A_c d (a_1-a_2)+\frac{1}{4\pi}A_s d (a_1+a_2)
\label{eq:critical_theory_appendix_1}
\end{align}
where $s$ is the tuning parameter.  $A_{c;\mu}$ and $A_{s;\mu}$ are probing field coupled to the total charge and the spin. 
 The $\mathcal M \mathcal C \mathcal T$ symmetry requires that $\vec{A}_s=0$. Physically the magnetic fields of the two layers are the same and exciton does not feel any net magnetic field.   From symmetry analysis, one can in principle include a term $\varphi^*(i\partial_t-A_{s;0})\varphi$.  This term is fine tuned to be zero because we are considering an interaction driven transition through $d/l_B$ at fixed exciton density.  On the other hand, one can also tune a FQSH to SF transition through tuning the displacement field, which corresponds to a chemical potential tuned transition.  The dynamical exponent for such a transition will be $z=2$ and is not our interest.

We then make a redefinition: $a_c=a_1+a_2, a_s=a_1-a_2$. The above actions change to:

\begin{align}
	\mathcal L&=|(\partial_\mu -ia_{s;\mu})\varphi|^2- s |\varphi|^2-g|\varphi|^4-\frac{3}{4\pi}a_c d a_s +\frac{1}{2\pi}A_c d a_s+\frac{1}{4\pi}A_s d a_c
\label{eq:critical_theory_appendix_2}
\end{align}

The redefinition of the gauge field changes the charge quantization rules.  The charge under $a_c$ is labeled as $q_c$ and the charge of $a_s$ is $q_s$. We have the charge transformation: $q_c=\frac{q_1+q_2}{2}$ and $q_s=\frac{q_1-q_2}{2}$, where $q_1,q_2$ are charges under $a_1,a_2$.  The elementary charge configuration is $(q_1,q_2)=(\pm 1,0), (0,\pm 1)$. Thus in terms of $(q_c,q_s)$, the elementary charge is $(q_c,q_s)=(\pm \frac{1}{2},\pm \frac{1}{2})$. In the FQSH phase with $\varphi$ gapped, one can check that for the excitation $(q_c,q_s)=(\pm \frac{1}{2}, \pm \frac{1}{2})$, it has statistics $\theta=\pm\frac{\pi}{3}$, physical charge $Q_c=\pm\frac{1}{3}$ and physical spin $Q_s=\pm\frac{1}{6}$. This is exactly the elementary anyon on layer 1 or layer 2  in the FQSH phase.  When $\varphi$ is condensed, $a_s$ is higgsed and we are left with the superfluid action $\mathcal L_{SF}=\frac{1}{4\pi}A_s d a_c$.  We can see that the vortex charge $q_v$ of the superfluid is $q_v=2q_c$. Then the elementary anyon with $(q_c,q_s)=\pm (\frac{1}{2},\frac{1}{2})$ in the FQSH phase now becomes the elementary vortex with $q_v=\pm 1$ in the superfluid phase. Of course now it costs infinite energy due to the coupling to the gapless gauge field $a_c$ which represents the Goldstone mode of the superfluid.  From this analysis one can see that the elementary anyon in the FQSH phase becomes the vortex of  the superfluid  in the exciton condensation (EC) phase. Its energy cost is infinite in the EC phase and finite in the FQSH phase. But it remains gapped across the phase transition. It is known that the elementary vortex become gapless and then condensed  in the usual superfluid to insulator transition.  Later in the dual theory we will see that at the XY* critical point what becomes gapless is a triple vortex, while the vortexes with $|q_v|=1,2$ remain gapped.

At the critical point, the topological property of the anyon does not matter. So we can integrate $a_c$ which simply locks $a_s=\frac{1}{3}A_s$.
 Then we reach  the final critical theory:

\begin{align}
	\mathcal L_c&=|(\partial_\mu -i\frac{1}{3}A_{s;\mu})\varphi|^2 -s |\varphi|^2-g|\varphi|^4+\frac{1}{6\pi} A_c d A_s
	\label{eq:critical_theory_appendix3}
\end{align}

When $s<0$, this is a superfluid phase of $A_s$. When $s>0$, we have the correct response of $\frac{1}{6\pi} A_c d A_s$ for the FQSH phase, though we have lost the information about the anyons by integrating $a_c$.

\subsection{Dual theory}

It is known that the XY critical theory such as in Eq.~\ref{eq:critical_theory_appendix3} has a dual theory. Here we derive the dual critical theory. We start from Eq.~\ref{eq:critical_theory_appendix_2} and apply the standard particle vortex duality for $\varphi$, then we obtain a dual critical theory as:

\begin{align}
	\mathcal L&=|(\partial_\mu -ib_{\mu})\tilde \varphi|^2-r |\tilde \varphi|^2-g |\tilde \varphi|^4-\frac{3}{4\pi}a_s d a_c+\frac{1}{2\pi} b d a_s+\frac{1}{2\pi}A_c d a_s+\frac{1}{4\pi}A_s d a_c
\end{align}

Integrating $a_s$, we locks $b=\frac{3}{2}a_c-A_c$. With a redefinition $a=\frac{1}{2} a_c-\frac{1}{3}A_c$, we get:

\begin{equation}
	\mathcal L=|(\partial_\mu -i3a_{\mu})\tilde \varphi|^2-r |\tilde \varphi|^2-g |\tilde \varphi|^4+\frac{1}{2\pi}A_s d a+\frac{1}{6\pi}A_s d A_c\notag\\
	\label{eq:dual_theory}
\end{equation}
which is exactly the dual theory of Eq.~\ref{eq:critical_theory_appendix3}. We choose the normalization of the gauge field $a$ so we have the usual coupling $\frac{1}{2\pi}A_s da$. When $\tilde \varphi$ is gapped in the $r>0$ side, this describes a superfluid phase of $A_s$. Here $\tilde \varphi$ here carries vortex charge $q_v=3$ of the superfluid phase. Therefore at the QCP the triple vortex instead of the elementary vortex of the superfluid becomes gapless. Condensation of this triple vortex kills the superfluid phase and leads to an insulator. The elementary vortex remains gapped across the QCP and will become the anyon in the FQHE insulator after the superfluid is gone.

\section{Extra-ordinary boundary criticality}

We start from the edge theory for the FQSH phase at $d>d_c$. For now let us consider the filling $(\nu_1,\nu_2)=(\frac{1}{3},-\frac{1}{3})$. From the K matrix $K=\begin{pmatrix} 3 & 0 \\ 0 & -3 \end{pmatrix}$ we can write down the effective action for the helical edge modes:
\begin{align}
	S&= \int dt dx \frac{3}{4\pi} \partial_t \varphi_1 \partial_x \varphi_1-\frac{3}{4\pi} \partial_t \varphi_2 \partial_x \varphi_2- \frac{3}{4\pi}\upsilon_F (\partial_x \varphi_1)^2-\frac{3}{4\pi}\upsilon_F(\partial_x \varphi_2)^2-g\frac{3}{2\pi}\upsilon_F (\partial_x \varphi_1)(\partial_x \varphi_2)
\end{align}
where $\varphi_1$ and $\varphi_2$ represent the edge mode in layer 1 and 2 respectively. The $g<0$ term is from inter-layer repulsion. Note in our current convention the density operators are $\rho_1=e\frac{1}{2\pi} \partial_x \varphi_1$ and $\rho_2=-e\frac{1}{2\pi} \partial_x \varphi_2$.

We have the commutation relations:

\begin{equation}
	[\varphi_1(x), \partial_y \varphi_1(y)]=i\frac{2\pi}{3} \delta(x-y)
\end{equation}
and
\begin{equation}
	[\varphi_2(x), \partial_y \varphi_2(y)]=-i\frac{2\pi}{3} \delta(x-y)
\end{equation}

Next we do a linear combination and define:

\begin{align}
\varphi_1(x)&=\frac{1}{\sqrt{3}}(\phi(x)+\theta(x))\notag \\ 
\varphi_2(x)&=\frac{1}{\sqrt{3}}(\phi(x)-\theta(x))\notag \\ 
\end{align}

so

\begin{equation}
	[\theta(x),\partial_y\varphi(y)]=i \pi\delta(x-y)
\end{equation}

This leads to the action:

\begin{equation}
 S=\int dt dx \frac{1}{\pi} \partial_t \phi \partial_x \theta-\frac{\tilde \upsilon_F}{2\pi} \big(K(\partial_x \theta)^2+\frac{1}{K}(\partial_x \phi)^2\big)
\end{equation}
where $\tilde \upsilon_F=\sqrt{1-g^2}\upsilon_F$ and $K=\frac{1-g}{1+g}$. So we have $K>1$.

We can also integrate $\phi$ to reach
\begin{equation}
	S=\int dt dx \frac{K}{2\pi  \tilde \upsilon_F}\big((\partial_t \theta)^2-\tilde \upsilon_F^2(\partial_x \theta)^2 \big)
\end{equation}

This is the standard action for the Luttinger liquid, but the operator mapping is different. Especially the fractional exciton $l=(1,-1)$ now corresponds to $\psi^\dagger\sim e^{i\varphi_R}e^{-i \varphi_L}=e^{i\frac{2}{\sqrt{3}}\theta(x)}$. It's scaling dimension is $[\psi]=\frac{1}{3K}<\frac{1}{3}$. We can make a redefinition $\tilde \theta=\frac{2}{\sqrt{3}}\theta$, so the action is:

\begin{equation}
	S_0=\int dt dx \frac{1}{2\pi  \tilde \upsilon_F \lambda}\big((\partial_t \tilde \theta)^2-\tilde \upsilon_F^2(\partial_x \tilde \theta)^2 \big)
\end{equation}
with $\lambda=\frac{4}{3K}$. In this convention $e^{i\tilde \theta}$ creates a fractional exciton with charge $1/3$ under $A_s$.

At the QCP, the boundary is described by the following action:

\begin{equation}
	S_{\text{boundary}}=\int dt dx \frac{1}{2\pi  \tilde \upsilon_F \lambda}\big((\partial_t \tilde \theta)^2-\tilde \upsilon_F^2(\partial_x \tilde \theta)^2 \big)-s \int dx dt (e^{i\tilde \theta} \varphi^*+e^{-i \tilde \theta}\varphi)
\end{equation}

Following Ref.~\onlinecite{metlitski2022boundary}, we obtain the renormalization flow equation:
\begin{align}
\frac{ds}{dl}&=(2-\Delta_{\varphi}-\frac{1}{4}\lambda)s \notag \\ 
\frac{d \lambda}{dl}&=-\pi^2 s^2 \lambda^2
\end{align}

Given that $\Delta_{\varphi}\approx 1.219$ and initially $\lambda<\frac{4}{3}$, we will have $s$ flows to infinite and $\lambda$ flows to zero when the RG flow $l$ approaches infinite.

The ordinary exciton order creation operator is $e^{i 3 \tilde \theta}$. In the FQSH phase its correlation function is:
\begin{equation}
	\langle e^{i3 \tilde \theta(x)}e^{-i 3 \tilde \theta(y)}\rangle \sim \frac{1}{|x-y|^{9\lambda}}
\end{equation}

Therefore at the QCP, because $\lambda$ flows to zero, the exponent of the above correlation function also flows to zero.  In practice, it should have a log singularity \cite{metlitski2022boundary} at the QCP:
\begin{equation}
	\langle e^{i3 \tilde \theta(x)}e^{-i 3 \tilde \theta(0)}\rangle \sim \frac{1}{(\log x)^q}\ , x \rightarrow \infty
\end{equation}

This is so-called extra-ordinary-log boundary critical behavior. One can see that the exciton order has an almost long-range order at the edge, despite that its correlation function has a large decaying exponent in the bulk.

The exciton current $A_s$ couples in the following way: $\partial_{\mu} \tilde \theta \rightarrow (\partial_\mu -i \frac{1}{3} A_{s;\mu})\tilde \theta$.  In the FQSH phase it is known the conductance under $A_s$ is $G=\frac{1}{9} \frac{1}{\lambda} \frac{e^2}{h}$\cite{giamarchi2003quantum}.  At the QCP, the conductance $G$ then flows to infinite. This is expected because the exciton has almost long-range order, so its transport should still be superfluid-like.

We also want to comment on the local density of states (DoS) of one layer probed by the scanning tunneling microscope (STM).  Consider the layer 1, the single electron creation operator is: $c_1^\dagger(x) \sim e^{i 3 \varphi_1(x)}=e^{i\sqrt{3}(\phi(x)+\theta(x))}$.  Then we expect the STM of the layer 1 has $\frac{dI}{dV}\sim V^{\alpha}$ with $\alpha=\frac{3}{2}(K+\frac{1}{K})-1$.   In the decoupled phase at large $d/l_b$ we have $\alpha \approx 2$ as $K\approx 1$. However, at the QCP, $\alpha$ goes to infinite in the extra-ordinary criticality.

\end{document}